\documentclass[longauth]{aa} 
\usepackage{graphicx}
\usepackage[T1]{fontenc}
\usepackage{amsmath}	
\usepackage{amssymb}	
\usepackage{times}

\usepackage{txfonts}
\usepackage{multirow}
\usepackage{longtable}
\usepackage{lscape}
\usepackage[]{natbib}
\usepackage{hyperref}
\bibpunct{(}{)}{;}{a}{}{,}
\makeatletter
\renewcommand*\aa@pageof{, page \thepage{} of \pageref*{LastPage}}
\makeatother

\begin{document}
   \title{The luminous host galaxy, faint supernova \\and rapid afterglow rebrightening of GRB\,100418A\thanks{This work makes use of data obtained at the following telescopes/observatories: VLT/Paranal (proposals 085.A-0009, 085.D-0773), GTC/ORM (proposals GTC74-10A, GTCMULTIPLE2B-17A), Keck/MK, Subaru/MK, 3.5m/CAHA, UKIRT/MK, WHT/ORM, RTT150/TUBITAK, {\it Spitzer}, PdBI/IRAM, WSRT/RO, Ryle/MRAO, and SMA/MK.}
   }

   \author{
   A. de Ugarte Postigo
          \inst{\ref{af:iaa},\ref{af:dark}}
          \and
   C.~C.~Th\"one \inst{\ref{af:iaa}}  
        \and
   K.~Bensch\inst{\ref{af:iaa}}  
	\and 
   A.~J.~van~der~Horst \inst{\ref{af:gwu},\ref{af:gwu2}} 
    \and
   D.~A.~Kann\inst{\ref{af:iaa}} 
   	\and
   	Z. Cano\inst{\ref{af:iaa}} 
   	\and
    L. Izzo\inst{\ref{af:iaa}} 
    \and
   	P.~Goldoni\inst{\ref{af:cnrs}} 
   	\and
   	S.~Mart\' in\inst{\ref{af:alma},\ref{af:esochile}}  
   	\and
   	R.~Filgas\inst{\ref{af:upraha}} 
   	\and
   	P.~Schady\inst{\ref{af:garching}} 
   	\and
    J.~Gorosabel\inst{\ref{af:upv},\ref{af:iker},\ref{af:iaa}}\textdagger 
    \and
   	I.~Bikmaev \inst{\ref{af:kazan},\ref{af:tatar}} 
   	\and
    M.~Bremer\inst{\ref{af:iram}} 
   	\and
   	R.~Burenin \inst{\ref{af:sri},\ref{af:mos}} 
   	\and
   	A.~J.~Castro-Tirado \inst{\ref{af:iaa}} 
   	\and
   	S.~Covino\inst{\ref{af:merate}}
   	\and
    J.~P.~U.~Fynbo \inst{\ref{af:dark},\ref{af:dawn}} 
   	\and
    D.~Garcia-Appadoo \inst{\ref{af:alma}} 
    \and
   	I.~de~Gregorio-Monsalvo \inst{\ref{af:alma},\ref{af:esochile}} 
    \and
    M.~Jel\'inek\inst{\ref{af:ondrejov}} 
   	\and
   	I.~Khamitov \inst{\ref{af:tubitak},\ref{af:kazan}} 
	\and
   	A.~Kamble\inst{\ref{af:cfa}} 
   	\and
   	C.~Kouveliotou\inst{\ref{af:gwu},\ref{af:gwu2}} 
   	\and
    T.~Kr\"uhler\inst{\ref{af:garching}} 
   	\and
    S.~Melnikov \inst{\ref{af:kazan},\ref{af:tatar}}
	\and
	M.~Nardini\inst{\ref{af:bicocca}} 
   	\and
	D. A. Perley \inst{\ref{af:liverpool}} 
	\and
    G.~Petitpas \inst{\ref{af:sma}} 
    \and      
    G.~Pooley \inst{\ref{af:cavendish}} 
	\and	
	A.~Rau\inst{\ref{af:garching}} 
   	\and
   	E.~Rol\inst{\ref{af:aus1},\ref{af:aus2}} 
	\and
	R.~S\'anchez-Ram\'irez\inst{\ref{af:iaps},\ref{af:iaa}} 
   	\and
    R.~L.C.~Starling\inst{\ref{af:leicester}} 
    \and
    N.~R.~Tanvir\inst{\ref{af:leicester}} 
    \and
    K.~Wiersema\inst{\ref{af:warwick}} 
    \and
    R.~A.~M.~J.~Wijers\inst{\ref{af:anton}}
    \and
    T.~Zafar\inst{\ref{af:aao}} 
          }

  \institute{
    Instituto de Astrof\' isica de Andaluc\' ia, Glorieta de la Astronom\'ia s/n, E-18008, Granada\\          \email{deugarte@iaa.es}\label{af:iaa}
    \and
	Dark Cosmology Centre, Niels Bohr Institute, Juliane Maries Vej 30, Copenhagen \O, 2100, Denmark\label{af:dark}
    \and
	The George Washington University, 725 21st street, NW, Washington D.C. 20052, U.S.A.\label{af:gwu}
	\and
	GWU/Astronomy, Physics and Statistics Institute of Sciences (APSIS)\label{af:gwu2}
	\and
    Astroparticules et Cosmologie, Universit\'e Paris Diderot, CNRS/IN2P3, CEA/Irfu, Observatoire de Paris, Sorbonne Paris Cité, 10 rue Alice Domon et Léonie Duquet, 75205, Paris Cedex 13, France\label{af:cnrs}
    \and
    Joint ALMA Observatory, Alonso de C\'ordova 3107, Vitacura 763 0355, Santiago, Chile\label{af:alma}
    \and
    European Southern Observatory, Alonso de C\'{o}rdova 3107, Vitacura, Casilla 19001, Santiago 19, Chile\label{af:esochile}
    \and
    Institute of Experimental and Applied Physics, Czech Technical University in Prague, Horska 3a/22, 12800 Prague, Czech Republic\label{af:upraha}
    \and
    Max-Planck-Institut f\"ur extraterrestrische Physik, Giessenbachstraße, 85748 Garching, Germany\label{af:garching}
    \and
	Unidad Asociada Grupo Ciencias Planetarias UPV/EHU-IAA/CSIC, Departamento de F\'isica Aplicada I, E.T.S. Ingenier\'ia, Universidad del Pa\'is-Vasco UPV/EHU, Alameda de Urquijo s/n, E-48013 Bilbao, Spain\label{af:upv}
    \and
    Ikerbasque,  Basque  Foundation  for  Science,  Alameda  de Urquijo 36-5, E-48008 Bilbao, Spain\label{af:iker}
    \and
	Kazan Federal University, Kremlevskaya Str., 18, 420008 Kazan, Russia\label{af:kazan}
    \and
	Academy of Sciences of Tatarstan, Bauman Str., 20, 420111 Kazan, Russia\label{af:tatar}
    \and
	Institut de Radioastronomie Millim\'etrique (IRAM), 300 rue de la Piscine, 38406 Saint Martin d’H\`eres, France\label{af:iram}
    \and
	Space Research Institute, Russian Academy of Sciences, Profsoyuznaya 84/32, 117997 Moscow, Russia\label{af:sri}
    \and
    National Research University Higher School of Economics, Moscow, Russia\label{af:mos}
    \and
    INAF-Osservatorio Astronomico di Brera, Via E. Bianchi 46, I-23807 Merate (LC), Italy\label{af:merate}
    \and
    Cosmic Dawn Center, Niels Bohr Institute, University of Copenhagen, Juliane Maries Vej 30, 2100 Copenhagen \O, Denmark\label{af:dawn}
    \and
    Astronomical Institute of the Czech Academy of Sciences, Frikova 298, CZ-251 65 Ondrejov, Czech Republic\label{af:ondrejov}
    \and
	TUBITAK National Observatory, Akdeniz Universitesi Yerleskesi, 07058, Antalya, Turkey\label{af:tubitak}
    \and
    Harvard-Smithsonian Center for Astrophysics, 60 Garden Street, Cambridge, MA 02138, USA\label{af:cfa}
    \and
    Universit\`a degli studi di Milano-Bicocca, Piazza della Scienza 3, 20126, Milano, Italy\label{af:bicocca}
    \and
    Astrophysics Research Institute, Liverpool John Moores University, IC2, Liverpool Science Park, 146 Brownlow Hill, Liverpool L3 5RF, UK\label{af:liverpool}
    \and
    Harvard-Smithsonian Center for Astrophysics, Submillimeter Array, 645 North A'ohoku Place, Hilo, HI 96720, USA\label{af:sma}
    \and
    Cavendish Laboratory, JJ Thomson Avenue, Cambridge CB3 0HE, UK\label{af:cavendish}
    \and
    Monash Centre for Astrophysics, Monash University, PO Box 27, Clayton, Victoria  3800, Australia\label{af:aus1}
    \and
    School of Physics and Astronomy, Monash University, PO Box 27, Clayton, Victoria 3800, Australia\label{af:aus2}
    \and
    INAF-Istituto Astrofisica e Planetologia Spaziali, Via Fosso Cavaliere 100, 00133 Rome, Italy\label{af:iaps}
    \and
    Department of Physics and Astronomy, University of Leicester, University Road, Leicester, LE1 7RH, UK\label{af:leicester}
    \and
    Department of Physics, University of Warwick, Coventry, CV4 7AL, UK\label{af:warwick}
    \and
    Anton Pannekoek Institute for Astronomy, University of Amsterdam, PO Box 94249, 1090 GE Amsterdam, The Netherlands\label{af:anton}
    \and
    Australian Astronomical Observatory, PO Box 915, North Ryde, NSW 1670, Australia\label{af:aao}
}

   \date{Received 13 June 2018; accepted 24 July 2018}

  \abstract
   {Long gamma-ray bursts (GRBs) give us the chance to study both their extreme physics and the star-forming galaxies in which they form.}
   {GRB\,100418A, at a redshift of $z=0.6239$, had a bright optical and radio afterglow, and a luminous star-forming host galaxy. This allowed us to study the radiation of the explosion as well as the interstellar medium of the host both in absorption and emission.}
   {We collected photometric data from radio to X-ray wavelengths to study the evolution of the afterglow and the contribution of a possible supernova (SN) and three X-shooter spectra obtained during the first 60 hr.}
   {The light curve shows a very fast optical rebrightening, with an amplitude of $\sim3$ magnitudes, starting 2.4 hr after the GRB onset. This cannot be explained by a standard external shock model and requires other contributions, such as late central-engine activity. Two weeks after the burst we detect an excess in the light curve consistent with a SN with peak absolute magnitude $\rm{M}_V=-18.5$ mag, among the faintest GRB-SNe detected to date. The host galaxy shows two components in emission, with velocities differing by 130 km s$^{-1}$, but otherwise having similar properties. While some absorption and emission components coincide, the absorbing gas spans much higher velocities, indicating the presence of gas beyond the star-forming regions. The host has a star formation rate of $\rm{SFR}=12.2$ M$_\odot$ yr$^{-1}$, a metallicity of $\rm{12+log(O/H)}=8.55$, and a mass of $1.6\times10^9$~M$_\odot$.}
   {GRB\,100418A is a member of a class of afterglow light curves which show a steep rebrightening in the optical during the first day, which cannot be explained by traditional models. Its very faint associated SN shows that GRB-SNe can have a larger dispersion in luminosities than previously seen. Furthermore, we have obtained a complete view of the host of GRB\,100418A owing to its spectrum, which contains a remarkable number of both emission and absorption lines.}


   \authorrunning{A. de Ugarte Postigo et al.}
   \titlerunning{GRB\,100418A host, supernova and rebrightening}
   \maketitle
%

\section{Introduction}

Long gamma-ray bursts (GRBs) are produced during the dramatic death of very massive stars in which the core collapses, forming a black hole, while material is ejected through narrow polar jets \citep{rho97}. In order to see the prominent gamma-ray emission, the observer's perspective needs to be within the opening angle of the jet. In the most accepted model \citep{pir99}, the gamma rays are produced during internal collisions of material moving at ultrarelativistic speeds within the jets. When the ejecta collide with the material that surrounds the star, they are decelerated and at the same time intense synchrotron radiation is emitted. This radiation can be observed at all wavelengths between radio and X-rays, and is known as the forward shock. This electromagnetic emission is often observed for several days before it fades away in optical and X-rays, but can occasionally be followed in radio for periods of weeks, months, or, in some exceptional cases, years. This synchrotron radiation is characterised by several power laws connected at characteristic frequencies \citep{sar98}. By studying their evolution we can determine some of the micro- and macro-physical parameters that drive the explosion. Occasionally the GRB afterglow can also have a reverse-shock component \citep{mes99,nak04}, which originates within the shocked material that bounces back into the jet after the encounter of the forward shock with the circumstellar material. This reverse shock can be quite luminous but is typically shorter lived than the forward shock.

In this basic afterglow model, the light curves are also expected to evolve according to simple broken power laws. However, when sufficient sampling is available, the light curves often deviate from a simple evolution, showing flares, bumps, and wiggles \citep[e.g. ][]{pal04,deu05} that require the addition of further parameters into the models. Energy injections either due to late central engine activity \citep{dai98,zha02} or to ejecta simultaneously emitted at different velocities \citep{ree98,sar00}, structured jets \citep{mes98}, patchy shells \citep{nak04b}, density fluctuations \citep{wan00,ram01}, or double jets \citep{ber03,fil11,kan18} are some examples of models that try to explain some of the deviations from the standard model.

Gamma-ray burst afterglows are some of the most luminous events known in astronomy and can be observed at virtually any redshift. The furthest spectroscopically confirmed GRB was measured at $z=8.2$ \citep{tan09,sal09}. Together with their clean synchrotron spectra, they are useful background light beacons for studying the absorption features of interstellar material within their host galaxies, the intergalactic medium, and any intervening material along their lines of sight. Furthermore, being produced by very massive stars, they are normally found in strongly star-forming (SF) galaxies, and within them, not far from their actual birthplaces, since their progenitor stars have rather short lives. This means that GRBs are great tools for stying SF galaxies throughout the Universe. Using GRBs to pinpoint these galaxies gives us the advantage of being able to find even the faintest dwarf SF galaxies \citep{hjo12,per16a}, which would be missed with regular surveys. As an additional advantage, the galaxies can be studied from within, in absorption, using the light of the afterglow, and later in emission, once the afterglow has faded away.

Since GRBs are the result of the collapse of massive stars, they naturally have an associated supernova (SN) \citep{gal98,hjo03}. The SNe associated with GRBs are normally classified as broad-lined Type Ic events, which are relativistic SNe with fast ejecta responsible for the broadening of the lines. They are also some of the most luminous events, similar in luminosity to Type Ia SNe, and only surpassed by the recently discovered superluminous SNe \citep{qui11}. Their very high kinetic energy is the reason why they have occasionally been called hypernovae \citep{pac98}. We now know that the characteristics of the SNe associated with GRBs do not significantly correlate with the energy released in gamma rays (\citealt{xu13,mel14}, but see also \citealt{lu18}). However, we have observed a range of ejecta velocities and supernova luminosities whose origin is not yet fully understood. There are even a few cases of long GRBs where an associated supernovae has not been observed \citep{fyn06,del06,mic16,tan17}, and a long ongoing debate is still trying to determine the reasons.

In this paper we present an extensive data set of observations of GRB\,100418A and its host galaxy at a redshift of $z=0.6239$, covering photometric observations from radio to X-rays. This data collection includes unpublished radio observations from GMRT and Ryle; mid-infrared imaging from {\it Spitzer}; optical and near-infrared (NIR) imaging from GTC, VLT, Keck, UKIRT, CAHA and WHT, as well as temporally resolved spectroscopic observations from the VLT/X-shooter. GRB\,100418A had a bright radio counterpart that was detected for two years after the burst. The optical light curve showed strong deviations from the simple broken power law expected from the fireball model. GRB\,100418A is also one of the few cases in which we have good enough data to study the host galaxy in detail, both in absorption and in emission. Finally, the data set that we collected allowed us to search for an associated supernova.

In Sect. \ref{obs} we present the observations and data analysis, Sect. \ref{ag} we discuss the characteristics of the afterglow evolution, Sect. \ref{sn} we explain our search for a supernova component, and Sect. \ref{host} we deal with the host galaxy. In Sect. \ref{con} we present the conclusions of our work. Throughout the paper we adopt a cosmology with a flat universe in which $H_0=67.8$ km s$^{-1}$ Mpc$^{-1}$, $\Omega_m=0.308$, and $\Omega_\lambda=0.692$ \citep{pla16}. We will follow the convention $F_\nu\propto t^{\alpha}\nu^{\beta}$ to describe the temporal and spectral evolution of the afterglow. Uncertainties are given at 68\% ($1\sigma$) confidence level for one parameter of interest unless stated otherwise, whereas upper limits are given at the $3\sigma$ confidence level.


\section{Observations}
\label{obs}
   
   \begin{figure*}
   \centering
   \includegraphics[width=17cm]{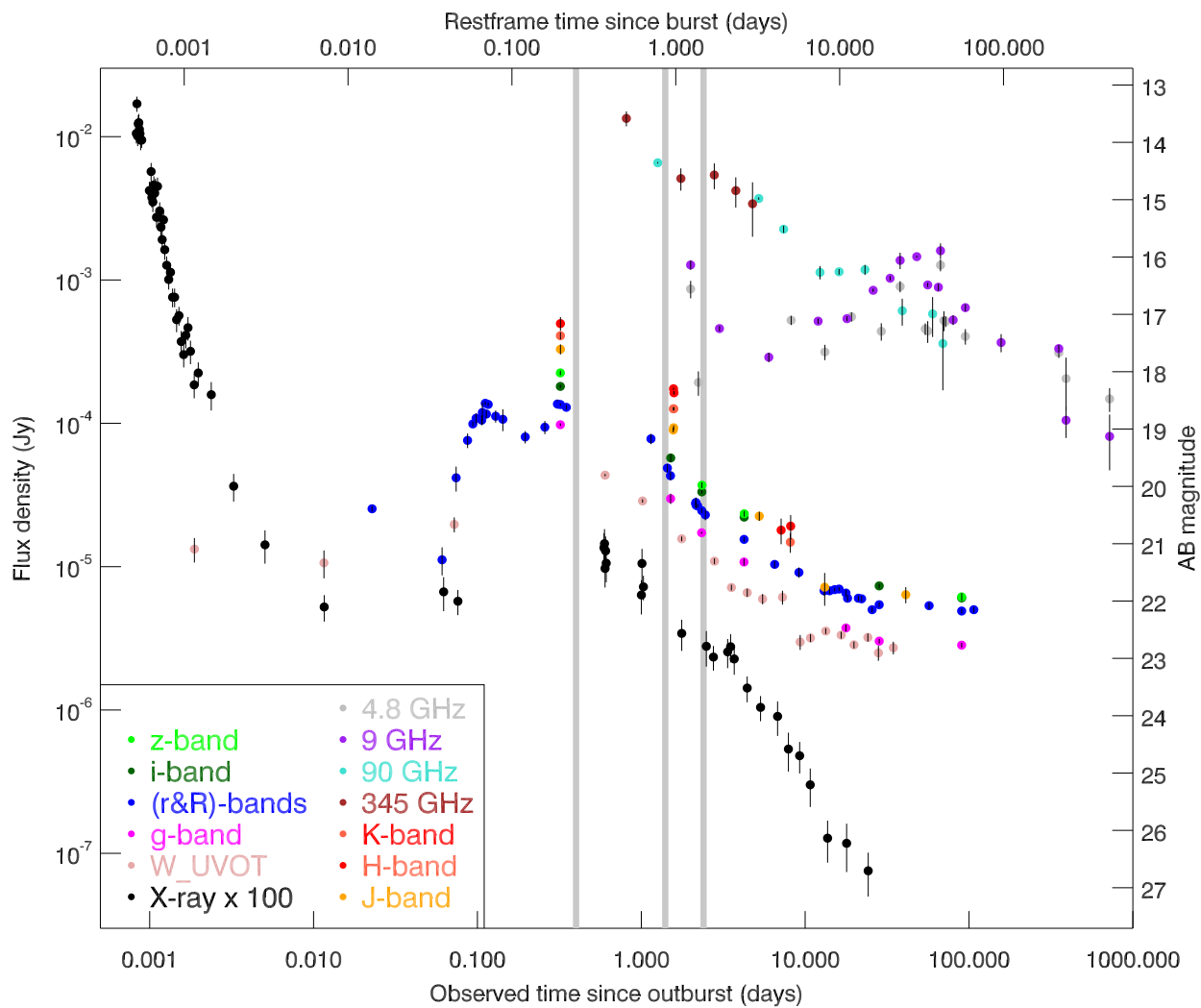}
      \caption{Light curves of GRB 100418A from X-ray to radio in all bands with significant coverage. Grey vertical lines indicate the times of the X-shooter spectral observations.
              }
         \label{Figlc}
   \end{figure*}

\subsection{\textit{Swift}  observations}

GRB\,100418A was detected by the Burst Alert Telescope (BAT) on-board the Neil Gehrels \textit{Swift} Observatory (\textit{Swift} hereafter) on 18 April 2010 at 21:10:08 UT \citep[][T$_0$ hereafter]{mar10}. The gamma-ray emission was characterised by two overlapping peaks,  with a duration of T$_{90}=7\pm1$ s in the 15-350 keV range (where T$_{90}$ is defined as the time during which 90\% of the photons are collected, from 5\% to 95\% of the total emission), and a hardness ratio (HR) of HR $=0.89$ \citep[][the HR is defined as the ratio of the fluence in the 50-100 keV band over the fluence in the 25-50 keV band]{ukw10}. This identifies GRB\,100418A as a long GRB according to the classical division \citep{kou93}, but also as a prototypical intermediate burst, according to the definition of \cite{hor10}, with a probability of 0.998 of belonging to this class. The prompt emission was also detected in the soft X-ray band by MAXI SSC \citep{ima16}.

Immediately after the trigger, \textit{Swift} slewed towards the field to observe with its narrow field instruments. The X-ray Telescope (XRT) began observations 79.1 s after T$_0$ and detected a bright, rapidly fading counterpart. The Ultraviolet/Optical Telescope (UVOT) started observing 87 s after T$_0$ and identified an ultraviolet/optical counterpart at a position consistent with the $\gamma$-ray and X-ray transient. These two instruments performed continued observations of the source for over a month, when it became too faint to be detected by the XRT and its ultraviolet/optical emission was dominated by the host galaxy. In this paper we make use of the X-ray and ultraviolet/optical data publicly available from \textit{Swift} \citep{eva07,eva09,marshall2011}.

\subsection{Near-infrared and Optical imaging}

Following the detection of the GRB and its afterglow by \textit{Swift}, observations were undertaken in the optical and NIR with ground-based telescopes \citep{pan10,fil10,upd10,ant10,cuc10,pea10,kle10,bik10}. For this work we gather imaging observations from observatories located across the world in order to obtain a dense coverage of the light-curve evolution.

Our earliest observations began 30 minutes after the burst at the 1.5 m Russian-Turkish Telescope (RTT150), located at TUBITAK National Observatory (Turkey) and were performed using an \textit{R$_C$} filter. This observatory followed the evolution of the source during the weeks that followed until the light was completely dominated by the host galaxy.

The 10.4~m Gran Telescopio Canarias (GTC) performed deep optical observations at four epochs between 3 and 90 days in search of the expected supernova component and to study the underlying host galaxy. Further optical images were obtained at different epochs with the European Southern Observatory's (ESO) Very Large Telescope equipped with FORS2 and X-shooter (as part of the spectroscopic observations), with the 10~m Keck telescope with LRIS, and with the United Kingdom Infrared Telescope (UKIRT) using the Wide Field Camera (WFCam). The host galaxy was observed when the afterglow and supernovae components where no longer significant, between 3 months and 7 years after the GRB. Our data set includes optical observations from OSIRIS at the 10.4~m GTC and LRIS at the 10~m Keck telescope, NIR data from Omega~2000 at the 3.5~m CAHA telescope and LIRIS at the 4.2~m WHT, and mid-infrared data from IRAC onboard the 0.85~m {\it Spitzer} space telescope. The summary of these host observations are shown in Table~\ref{table:hostmags}.

Data reduction was performed in a standard fashion using IRAF routines \citep{Tody1993ASPC}. Photometric calibration was obtained from field stars present in the SDSS catalogue, 2MASS for the NIR data, and tabulated zero points for {\it Spitzer}. All the photometric observations are presented in Table~\ref{table:vis}.

Last, we also include observations performed with the 8.2~m SUBARU telescope using FOCAS, data already published by \cite{nii12}, as well as early observations of the afterglow by BOOTES-2 \citep{jel16}. 

Throughout the paper we use the equatorial coordinates of the afterglow (J2000) determined from the PdBI millimetre data, with an uncertainty of 0\farcs15: R.A. = 17:05:27.091, Dec. = +11:27:42.29. Using these coordinates we can estimate the Galactic extinction along the line-of-sight to be $E(B-V)=0.065$ mag \citep{sch11}.

\addtocounter{table}{1}

\subsection{Submillimetre, millimetre and radio observations}

 Submillimetre (submm) data were obtained from the Submillimeter Array (SMA), in Hawaii (USA), starting 16 hr after the burst. For this first observation we used 7 out of the 8 SMA antennas, under good weather conditions, with zenith opacities at 225~GHz of $\tau\sim0.06$ (precipitable water vapour, PWV~$\sim1$~mm). Titan and Neptune were used as flux calibrators and 3C454.3 as bandpass calibrator. Atmospheric gain was corrected by observing the nearby quasar J1751+096 every 15 min. With these data we discovered a very bright counterpart \citep{martin10} at a flux of $13.40\pm1.60$ mJy. Observations continued over the following four nights, tracking the evolution of the afterglow until it became undetectable on 23 April.

Following the submm detection we observed the burst from Plateau de Bure Observatory (PdBI) of the Institut de Radioastronomie Millim\'etrique (IRAM). The observations were performed in the 3mm band and cover the temporal range between 1.3 and 70 days after the GRB. Both the SMA and the PdBI data were already published in the compilation of \cite{deu12a}.

Observations at 15~GHz were performed by the Ryle telescope. However, these observations are affected by blending with a bright nearby source, NVSS J170527.66 +112706.9. The afterglow is well detected between day 7 and day 10, but the rest of the epochs are probably dominated by the contribution of the contaminating source.

The afterglow of GRB\,100418A was detected with the Westerbork Synthesis Radio Telescope (WSRT) at 4.8~GHz two days after the burst at a flux density of about 200~$\mu$Jy. WSRT subsequently followed up the evolution of this afterglow at 4.8~GHz with a bandwidth of $8\times20$ MHz and the bright radio source 3C286 as a primary flux calibrator. The observations have been analysed using the Astronomical Image Processing Software (AIPS)\footnote{\url{http://www.aips.nrao.edu/}} and the Multichannel Image Reconstruction Image Analysis and Display \citep[MIRIAD,][]{sau95} software package.

The details of our observations and the results are shown in Table~\ref{table:radio}. In addition to this, we use the available literature data obtained from the Very Large Array (VLA), Australia Telescope compact Array (ATCA) and Very Long Baseline Array (VLBA) at 4-9 GHz for our analysis \citep{moin2013}.

\begin{table}
\caption{Radio observations}             
\label{table:radio}      
\centering                          
\begin{tabular}{c c c c}        
\hline\hline                 
T-T$_0$ 		& Observatory & Frequency 	& Flux 			    \\    
(days)     		&           &     (GHz)     & (mJy) 			\\
\hline                        
2.2315  		& WSRT 		& 4.8			& 0.193$\pm$0.037 	\\ 
8.2150  		& WSRT		& 4.8			& 0.524$\pm$0.039 	\\
13.2010		    & WSRT		& 4.8			& 0.314$\pm$0.038 	\\
19.1850		    & WSRT		& 4.8			& 0.554$\pm$0.041 	\\
29.1575		    & WSRT		& 4.8			& 0.438$\pm$0.061 	\\
54.0895		    & WSRT		& 4.8			& 0.456$\pm$0.041 	\\
\hline
4.7669		    & Ryle		&15.0		    & 0.90$\pm$0.12\tablefootmark{a}	\\
7.8410		    & Ryle		&15.0		    & 2.70$\pm$0.12\tablefootmark{a}	\\
8.8539		    & Ryle		&15.0	    	& 1.80$\pm$0.11\tablefootmark{a}	\\
9.7480		    & Ryle		&15.0	    	& 1.70$\pm$0.14\tablefootmark{a}	\\
13.8820		    & Ryle		&15.0   		& 1.10$\pm$0.18\tablefootmark{a}	\\
23.7130		    & Ryle		&15.0	    	& 0.70$\pm$0.11\tablefootmark{a}	\\
43.7400		    & Ryle		&15.0	    	& 1.50$\pm$0.12\tablefootmark{a}	\\
\hline
1.2603		    & PdBI		& 103.1	    	& 6.57$\pm$0.07	\\
5.2114		    & PdBI		& 86.7	    	& 3.70$\pm$0.07	\\
7.3930		    & PdBI		& 86.7	    	& 2.26$\pm$0.13	\\
12.3284		    & PdBI		& 105.7	    	& 1.13$\pm$0.12	\\
16.1079		    & PdBI		& 86.7	    	& 1.14$\pm$0.05	\\
23.1766		    & PdBI		& 86.7	    	& 1.18$\pm$0.09	\\
39.1444		    & PdBI		& 102.6	    	& 0.61$\pm$0.13	\\
59.9582		    & PdBI		& 86.7	    	& 0.58$\pm$0.18	\\
69.2371		    & PdBI		& 99.5	    	& 0.36$\pm$0.19	\\
\hline
0.8110   		& SMA 		& 350   		& 13.4$\pm$1.6 	    \\ 
1.7448   		& SMA 		& 350 	    	& 5.1$\pm$0.9 		\\ 
2.7891   		& SMA 		& 350 	    	& 5.4$\pm$1.1 		\\ 
3.7744   		& SMA 		& 350   		& 4.2$\pm$1.0 		\\ 
4.7754   		& SMA 		& 350    		& 3.4$\pm$1.4 		\\

\hline                                   
\end{tabular}
\tablefoot{
\tablefoottext{a}{Blended with the nearby source NVSS J170527.66 +112706.9.}
}
\end{table}

\subsection{Spectroscopy}

As soon as the field became observable from Paranal Observatory (Chile), we obtained follow-up spectroscopic observations of the afterglow of GRB\,100418A using X-shooter \citep{ant10}. X-shooter \citep{ver11} is a triple spectrograph capable of observing simultaneously a complete spectrum between 3000 and 24800 {\AA} at intermediate resolution. 
In our case, we used slit widths of of 1$^{\prime\prime}$, 0\farcs9 and 0\farcs9 in the UVB, VIS and NIR arms, respectively, with a $2\times1$ binning (binning in the spectral but not in the spatial direction), which delivers a resolution of 5100, 8800 and 5300 in each arm.

Thanks to the peculiar evolution of the afterglow, we were able to obtain a high signal-to-noise (S/N) ratio starting 8.37 hours after the burst onset, when the afterglow was near its maximum brightness, at $R_C=18.1\pm0.1$ \citep{ant10}. This spectrum presents abundant absorption and emission features over the entire spectral range of X-shooter at a redshift of $z = 0.6239\pm0.0002$, the redshift of the GRB. 

\begin{table}
\caption{Log of spectroscopic observations obtained with X-shooter at VLT.}             
\label{table:spec}      
\centering                          
\begin{tabular}{c c c c}        
\hline\hline                 
T-T$_0$ 			& Arm	 	& Exposure		& Average	\\    
(days)     			&                        	& (s)		      		&  S/N		\\
\hline                        
				& UVB		& 4$\times$1200	& 35		\\
0.375			& VIS		& 4$\times$1200	& 43		\\ 
				& NIR		& 8$\times$600	& 17		\\
\hline                                   
				& UVB		& 4$\times$1200	& 19		\\
1.470			& VIS		& 4$\times$1200	& 22		\\
				& NIR		& 8$\times$600	& 6.9		\\
\hline                                   
				& UVB		& 4$\times$1200	& 10		\\
2.487			& VIS		& 4$\times$1200	& 13		\\
				& NIR		& 8$\times$600	& 3.1		\\
\hline                                   
\end{tabular}
\end{table}

The sustained brightness of the afterglow allowed further observations with X-shooter over the following nights. In total, three epochs were obtained at 0.4, 1.5 and 2.5 days after the burst, with S/N ratios of 43, 22 and 13 per spectral bin in the VIS arm, respectively \citep{deu11b}. 
Each one of the three observations consisted of 4 different
exposures of 1200 s each; for the NIR arm each exposure was split in two individual ones of 600 s each.
The exposures were taken using the nodding along the slit 
technique with an offset of 5$^{\prime\prime}$ between exposures in a standard ABBA sequence. 
The log of spectroscopic observations is displayed in Table~\ref{table:spec}.

We processed the spectra using the X-shooter data reduction pipeline \citep{gol06,mod10} and further scripts, as described by \cite{sel18}. 
To perform flux calibration we extracted a spectrum from
a staring observation of the flux standard GD153
\citep{boh95} taken the same night for
each observation. Each of the spectra were matched to the multiband photometry interpolated to the mean time of each spectrum. For each epoch we performed two different extractions. First we use optimal extraction, which is best for the study of the afterglow. Then  we used one with a fixed aperture that allowed us to cover the full extension of the host galaxy. This second extraction provides us with better measurements for the study of the host.
The spectra are publicly available through the GRBSpec database\footnote{\url{http://grbspec.iaa.es}} \citep{deu14}.


\section{Afterglow analysis}
\label{ag}
\subsection{Light curves}

The first five minutes of the X-ray light curve showed a very sharp decline (with a temporal index of $\alpha=-4.23^{+0.17}_{-0.16}$, \citealt{pag10}), probably related to the tail of the prompt burst emission. This was followed by a flattening and subsequent increase in flux density, reaching a maximum around 0.5 days after the burst. After this, the light curve decayed with a steepening slope until it faded beyond the sensitivity limit of XRT 25 days after the burst.

The optical light curve was characterised by an early shallow decay, reaching a minimum at around 1.5 hr after the burst, after which there was a very steep increase in brightness of a factor of ten within one hour. After this step in the light curve the brightness was sustained, with perhaps a shallow decay until around one day, when the light curve steepened and decayed for the following 2 weeks, when it became dominated by the contribution of the host galaxy.

\begin{figure}
   \centering
   \includegraphics[width=8cm]{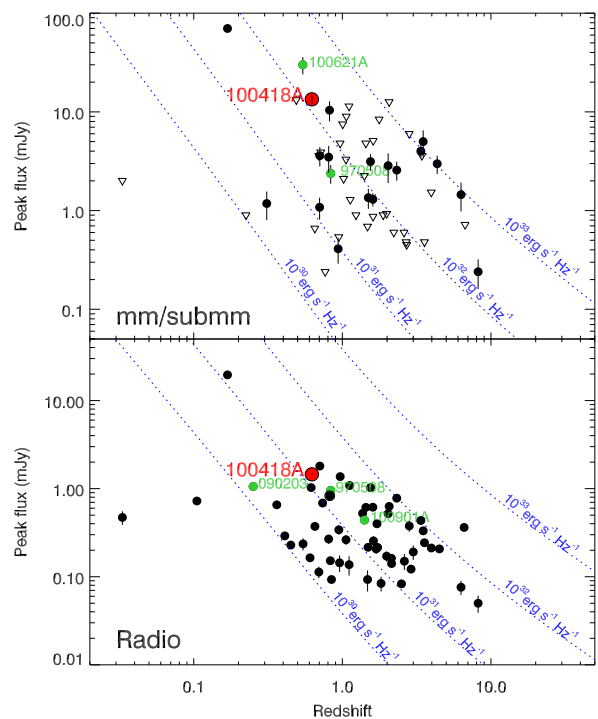}
      \caption{Peak flux densities of GRB\,100418A as compared to the mm/submm and radio samples from \cite{deu12a} and \cite{cha12}. In blue we indicate lines of equal luminosity, with the most luminous events being found in the top right region of the plots. We have highlighted in green other GRBs where strong rebrightenings have been observed (see Sect.~\ref{bbag}).
              }
         \label{FigRadio}
   \end{figure}

The submm and mm light curves at 350 GHz and 90 GHz evolved both in a similar way. There was a fast decay between the first observation obtained 0.8 days after the burst and day 1.5, after which the light curve first flattened for a day and then resumed its decay until day seven, when a second plateau stabilised the flux density until day 23. The light curve then decays and becomes too faint to be detected two months after the burst. At the time of discovery, this was the second brightest mm/submm counterpart detected (after GRB\,030329, which peaked at $\sim$70 mJy; \citealt{she03,res05}), and is currently still the 4th (after GRB\,171205A, \citealt{deu17} and GRB\,100621A, \citealt{gre13}), see Fig.~\ref{FigRadio}.

At 4.8 GHz, the light curve rose during the two months of observations that were performed with the WSRT. However, oscillations in the flux are apparent during the first 3 weeks, probably due to scintillation, as is commonly seen at these frequencies \citep{goo97}. Observations from the literature \citep{moin2013} also showed a peak in radio that was reached at around two months, followed by a steady decay.

Figure~\ref{FigRadio} shows the brightness of GRB\,100418A as compared to two samples of mm/submm \citep{deu12a} and radio afterglows \citep{cha12}. Although it is among the brightest events ever observed, the actual luminosity is not extraordinary, with a group of GRBs reaching luminosities almost an order of magnitude more than this event.

\subsection{Intrinsic dust extinction along the GRB line of sight}\label{AG:ext}

\begin{table}
\caption{Determination of intrinsic extinction laws through SED fitting. Fits shown here were performed with a single power-law spectrum.}             
\label{table:SEDfit}      
\begin{small}
\centering                          
\begin{tabular}{c c c c c}        
\hline\hline                 
Ext.  &	A$_V$			& 	N$_{\rm H}$				&	$\beta$					&	$\chi^2/d.o.f.$	\\ 
law			&	(mag)			&	$10^{22} cm^{-2}$			&							&				\\
\hline\hline \vspace{1mm}
SMC			&	$0.086\pm0.039$	&	$0.242_{-0.133}^{+0.170}$	&	$-1.061_{-0.023}^{+0.024}	$	&	90.25/95		\\\vspace{1mm}
LMC			&	$0.091\pm0.041$	&	$0.243_{-0.134}^{+0.171}$	&	$-1.062_{-0.023}^{+0.024}	$	&	90.32/95		\\\vspace{1mm}
MW			&	$0.092\pm0.042$	&	$0.242_{-0.133}^{+0.171}$	&	$-1.062_{-0.023}^{+0.024}	$	&	90.50/95		\\
\hline\hline
\end{tabular}
\end{small}
\end{table}

   \begin{figure}[ht!]
   \centering
   \includegraphics[width=8cm]{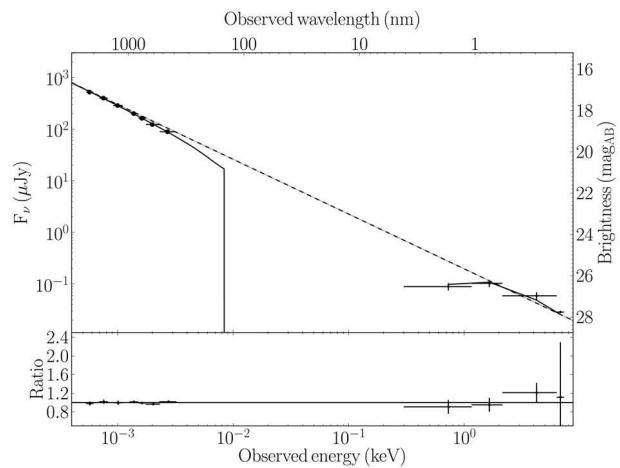}
      \caption{Fit of the SED at 0.4 days from NIR to X-rays with a single power law and an SMC extinction law.
              }
         \label{FigSEDfit}
   \end{figure}

We performed fits of the spectral energy distribution from NIR to X-rays at 0.4 days once the photometric data had been corrected for the foreground extinction stemming from our own Galaxy \citep{sch11}. We tested three intrinsic dust models (Small Magellanic Cloud -- SMC, Large Magellanic Cloud -- LMC, and Milky Way -- MW), and underlying simple and broken power-law spectra. The statistically preferred fit was obtained with an SMC extinction law and single power law, although at this redshift all the other dust laws are not ruled out and give comparable results \citep{kan06,sta07,zaf11,sch12}. If a broken power law is forced, the break is found at energies larger than X-rays, confirming the existence of a single power-law spectrum between optical and X-rays. We performed $\chi^2$ statistics for optical data and C-stat statistics for X-ray data. All errors are at 90\% confidence. Our best fit implies an SMC extinction law with an $A_V=0.086\pm0.039$ mag.

Several other works have already performed some kind of fit to estimate the line-of-sight extinction towards GRB\,100418A. The values that we derived are consistent with the measurement recently performed by \citet{zaf18}, who determined that the NIR to X-rays SED could be well explained by using a single power law and a featureless extinction curve with $R_V=2.42_{-0.10}^{+0.08}$ and $A_V=0.12\pm0.03$ mag. This is also in accordance with the findings of \citet{marshall2011} who describe the data with a single power law, and an SMC extinction of $A_V=0.056\pm0.048$ mag. We also note that our fits are inconsistent with those performed by \citet{jap15}, who found their best solution with a broken power law and an $A_V=0.20_{-0.02}^{+0.03}$ mag. Each of these works used slightly different methods and calibrations.

\subsection{Broadband afterglow evolution within the fireball model}
\label{bbag}

GRB\,100418A is one of the members of a growing class of GRBs that exhibit a fast rebrightening in their optical light curves within the first day after the gamma-ray trigger. Previous studies of this particular GRB have suggested different explanations for this feature, 
such as continuous injection of energy into the blast wave moving through an ambient medium, or an off-axis jet viewing angle \citep{marshall2011,laskar2015}. The injection of energy could be due to ejecta with a wide range of Lorentz factors or continued activity of the central engine, and previous multiwavelength studies have indicated that in the case of GRB\,100418A the continued activity of the central engine is favoured because the derived physical parameters for the other energy injection model are not plausible \citep{marshall2011,moin2013,laskar2015}. 
Previous studies, and in particular \citet{laskar2015}, performed a detailed broadband modelling of the evolution of this event, which we will not attempt to repeat here. However, they did not have or did not model the detailed optical to NIR range light curves between $\sim0.1$ and $\sim1$ day after the gamma-ray trigger. The light curve shows that after the very steep rise, there is a plateau that lasts a large fraction of the first day. Here we will focus on the steep rise, plateau, and subsequent ``normal'' afterglow decay, and possible explanations for the observed behaviour.

First of all, we determine the spectral energy distributions (SEDs) from NIR to X-ray frequencies for several epochs with good spectral coverage and for which the host galaxy contribution had been removed. This analysis was done at 0.60, 1.77 and 2.56 days, in a similar fashion to what was done in the previous section. We loaded the XRT spectra and NIR/optical magnitudes into \texttt{XSPEC}, and fit the SEDs with single and broken power-law functions. To account for optical extinction we used MW, SMC and LMC extinction models, and determined that there were no large differences with the different extinction laws due to the small amount of extinction and the low redshift. The main result is that the spectrum can be fit well with a single power law from NIR to X-ray frequencies at all epochs, and that a spectral break does not improve the fit. The spectral index is $\beta=-1.02\pm0.03$, $\beta=-1.05\pm0.06$ and $\beta=-1.00\pm0.05$ for each of the three epochs, respectively, indicating that there is no significant spectral evolution; most notably, that there is no spectral evolution from the plateau phase to the ``normal'' decay phase. Unfortunately there is insufficient spectral coverage during the light-curve rise, and it is thus not possible to check for any spectral evolution at early times. For further considerations we adopt the spectral index of the 0.6 day epoch, i.e. $\beta=-1.02\pm0.03$, as the spectral index for all three light-curve phases.

To compare our light curve with a standard fireball model, we look at the evolution after day 1, when the behaviour corresponds to that of a typical afterglow. Since the optical emission is affected by a strong contribution from the host galaxy and a possible supernova (see Sections~\ref{sn} and \ref{host}), we use the X-rays to perform a light curve evolution fit, as it is expected to contain only afterglow contribution. We find that beyond 1 day the X-ray light curve is well described by a broken power-law fit with a break time at $t_b=4.2\pm1.1$ day and temporal slope indices of $\alpha_1=-0.98\pm0.10$ and $\alpha_2=-1.91\pm0.16$. This break in the light curve is consistent with the jet break that we would expect to find in a collimated afterglow emission. Given that the spectral fits show that optical and X-ray data were within the same spectral segment, the X-ray afterglow light curve and the optical afterglow are expected to evolve in the same way. Indeed, in Sect.~\ref{sn} we show that the optical and X-ray afterglow evolution are consistent after day 2.

To further constrain the physics of the blast wave, we use the typical relations for the forward-shock model between the spectral index, the temporal indices, and the power-law index $p$ of the electron energy distribution (also known as the `closure relations', \citealt{Gao2013NewARG}). 
We find that the evolution is consistent with that of an afterglow expanding in the interstellar medium, where the cooling break would lie redwards of the optical/NIR data. This is consistent with what has been previously proposed for this event \citep{marshall2011,moin2013,laskar2015}. In such a scenario we can use the different spectral and temporal indices to derive the value of $p$: From the post-break slope we obtain that $p=-\alpha_2=1.91\pm0.16$, from the pre-break decay $p=(2-4\alpha_1)/3=1.97\pm0.14$, and from the spectral slope $p=-2\beta=2.04\pm0.06$. Combining them all, we can derive a best estimate of $p=2.02\pm0.05$. This value is not far from the value of $p=2.14$ estimated by \citet{laskar2015} from broadband modelling and within typical values for GRB afterglows \citep[e.g.][]{starling2008,curran2010,kan18}.

Given the value for $p$, we derive the rate of energy injection into the forward shock during the plateau phase. We adopt the convention from \citet{vaneerten2014} that the injected luminosity is proportional to $t_{in}^q$, with $t_{in}$ the energy injection timescale, i.e. the injected energy is proportional to $t_{in}^{1+q}$. We fit the optical-to-NIR light curves during the plateau (for which we assume a temporal slope of $0.07\pm0.07$), from 0.1 to 0.8~days after the burst, and find $q=-0.08\pm0.09$, i.e. energy injection at a constant rate. We note that this energy injection rate was not derived by earlier studies, because they did not have or did not model this part of the light curves. 

The steep rise between 0.05 and 0.1~days is much harder to interpret within the standard forward-shock model. The temporal slope is $\alpha_{rise}=5.4\pm1.1$ (using as origin the GRB trigger time, T$_0$), and if one would interpret this as an energy injection, this would lead to $q=5.3\pm1.1$, which seems to be quite extreme but not unfeasible. The energy increase in this time interval is a factor of $24_{-9}^{+15}$, while it is a factor of $6.6\pm0.4$ during the plateau phase. If one chooses to interpret the energy injection as ejecta with a wide range of Lorentz factors, the shallow decay phase can be interpreted in this way. In this theoretical framework, it is assumed that the amount of mass in the ejecta with a Lorentz factor greater than $\gamma$ is a power-law function of $\gamma$ with power-law index $s$ \citep[e.g. ][]{zhang2006}. The value we derived for $q$ results in $s=4.5\pm0.5$ for a fireball expanding in a constant density medium, which is reasonable, but the value for $s$ is very poorly constrained for a wind density medium ($\rho\propto r^{-2}$). Seeing these numbers, one would have a preference for the constant density scenario, as the s value is more reasonable. However, in the steep rise phase, the value for $s$ is unphysical (in both environment density scenarios), and therefore the latter phase cannot be interpreted in this way. Taking all of this into account, we agree with the general conclusion of previous studies that a model of continued activity of the central engine is favoured over ejecta with a wide range of Lorentz factors \citep{marshall2011,moin2013,laskar2015}, but our detailed analysis of the light curves and spectra leads to different conclusions on the rate of energy injection and details of the physics of the forward shock.

\begin{figure}[ht!]
   \begin{centering}
   \includegraphics[width=\columnwidth]{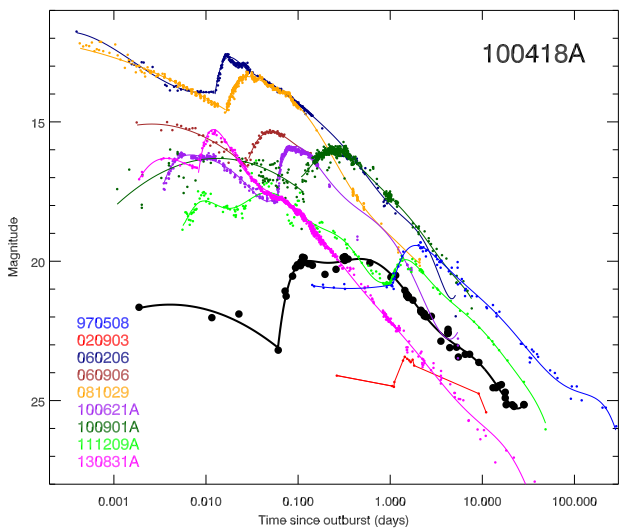}
      \caption{Comparison of the optical light curve of GRB\,100418A (in black) with other GRB afterglows showing steep rebrightenings. All the light curves have been shifted to a redshift of $z=1$ to allow for a direct comparison. When enough data are available the light curves have been fitted with two spline curves, one before the rebrightening and a second one after. When possible the contribution of the host galaxy has been subtracted.
              }
         \label{FigLcComp}
   \end{centering}
   \end{figure}

In Fig.~\ref{FigLcComp} we compare the light curve of GRB\,100418A with those of other GRB light curves with a similar steep re-brightening (from \citealt{kan06,kan10,kan17,kan18} and Kann et al. in prep.). In Table~\ref{table:lcparameters} we compare some light curve parameters of GRB\,100418A with those of these other GRBs. 

\begin{table}
\caption{Comparison of the characteristics of light curve rebrightenings seen for other GRB afterglows. The values have been calculated after shifting all the light curves to $z=1$.}
\label{table:lcparameters}      
\begin{center}                          
\begin{tabular}{c c c c}        
\hline\hline                 
GRB     &  $\alpha_{rise}$  & $\Delta mag$  & $t_{rise}$ (day)  \\ \hline\hline  
100418A &  $5.4\pm1.1$      & $3.4\pm0.5$   & $0.062\pm0.005$       \\
\hline
970508  &  $3.93\pm0.12$    & $1.41\pm0.12$ & $1.08\pm0.03$         \\
020903  &  $2.97\pm0.10$    & $1.02\pm0.10$ & $1.10\pm0.02$         \\
060206  &  $5.43\pm0.20$    & $1.27\pm0.12$ & $0.0125\pm0.0010$     \\
060906  &  $2.02\pm0.28$    & $1.28\pm0.40$ & $0.029\pm0.002$       \\
081029  &  $4.33\pm0.21$    & $1.19\pm0.14$ & $0.0165\pm0.0010$     \\
100621A &  $15.8\pm1.32$    & $1.96\pm0.16$ & $0.058\pm0.002$       \\
100901A &  $4.09\pm0.55$    & $1.45\pm0.25$ & $0.110\pm0.005$       \\
111209A &  $3.74\pm0.18$    & $0.90\pm0.05$ & $1.007\pm0.010$       \\
130831A &  $2.83\pm0.11$    & $1.41\pm0.14$ & $0.0082\pm0.0003$     \\
\hline\hline
\end{tabular}
\end{center}
\end{table}

These steps have been detected for a number of GRB afterglows. They tend to happen during the first day and have an optical luminosity increase ranging between 1 and 3 magnitudes. They are characterised by a very steep rise slope with typical values of $3<\alpha<5$ but that can reach values of $\alpha\sim15$. When enough multiband data has been available, a colour evolution has been evident (e.g. GRB\,081029, \citealt{nar11}; GRB\,100621A, \citealt{gre13}; GRB\,111209A, \citealt{kan18}). The afterglow is normally bluer before the step. The temporal slope is steeper in the redder bands, and the amplitude of the step is also larger in the redder bands. This is consistent with the onset of a different spectral component, however the actual nature of these bumps is not yet clear.

\section{The faint supernova associated to GRB\,100418A}
\label{sn}

The search for a supernova component in the light curve of GRB\,100418A is strongly affected by the light of the host galaxy, which is more than a magnitude brighter than what we would expect for a contribution similar to SN\,1998bw at this redshift. To get reliable photometry at the time when one could expect a SN to emerge, image subtraction using a host galaxy template is mandatory.

We used our deep GTC/OSIRIS images to obtain image-subtracted magnitudes of the optical emission associated with GRB\,100418A, using the final epoch in each filter as a template.  Image subtraction was performed using an adaptation of the original ISIS programme \citep{ala98,ala00} that was developed for the Hubble Space Telescope SN surveys by \cite{str04}, and used in previous GRB-SN studies \citep{can11,can14,can17}.  A key advantage of this code is the option to specify a set of stamps that the programme uses when it calculates the point-spread function in each image.  The image-subtraction technique was then optimised by varying the kernel mesh size and measuring the standard deviation ($\sigma$) of the background counts in a nearby region in the image (where images with lower $\sigma$ values indicate that they are a better subtracted image).  As a self-consistency check, we compared the magnitudes of the optical transient (OT, i.e. the combination of afterglow and SN emission) against those found by performing photometry on the un-subtracted images, converting the magnitudes into fluxes, and then mathematically subtracting the host flux. In images where the OT was detected in the subtracted image, excellent agreement was obtained with both methods, showing that the image-subtraction technique was well-optimised.  Using this method, an OT was seen at the position of the GRB in the subtracted images, 28.3 d after the burst, or 17.4 d in rest frame, when the supernovae would be expected to dominate the emission with photometric measurements corresponding to $g'=25.64\pm0.27$ mag, $r'=24.81\pm0.12$ mag and $i'=23.79\pm0.08$ mag.

\begin{figure}
   \centering
   \includegraphics[width=\columnwidth]{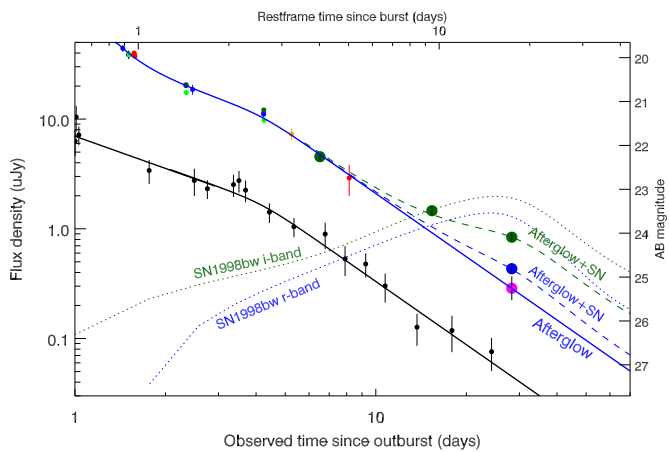}
      \caption{Late light curve of GRB\,100418A in X-rays and optical, where the contribution of the host galaxy has been removed. All the optical bands have been offset to the $r'$-band for clarity. Only data with signal-to-noise ratio above 3, after host subtraction are plotted. At late times image subtraction techniques (thick dots) show that there is an excess of emission in the $r'$ and $i'$ bands, corresponding to a faint supernova. Solid lines correspond to the afterglow contribution in X-rays and optical. The dashed lines show the combined contribution of afterglow and supernova (for which we used SN\,1998bw templates) scaled to match the late photometry. The dotted lines show the expected light curves for SN\,1998bw at the redshift of GRB\,100418A. Colour coding is equivalent to that of Fig.~\ref{Figlc}.}
         \label{fig:sn}
   \end{figure}

The magnitudes calculated above are a contribution of both afterglow and supernova. To estimate the actual luminosity of the SN, we need to determine the fraction of the light that comes from the afterglow at this time. To do this, we use the afterglow fit described in the previous section.
In Fig.~\ref{fig:sn} we have used the fit of the X-ray data (in black) and applied it to the optical light curves, which are well reproduced by this fit beyond day 2. At late times we see that the $g'$-band emission measured through image subtraction (big dots) is consistent with being due to only afterglow, but that both $r'$ and $i'$ bands show a clear excess due to a supernova component. Comparing to the expected flux for a supernova with the same luminosity as SN\,1998bw (corrected for a redshift of $z=0.6239$, dotted lines), we can already see that the supernova that we measure is significantly fainter than the usual standard for GRBs. We perform a similar analysis with the VLT/FORS data in the $I_C$ band on days 6 and 15, compared with the same filter on Keck/LIRIS. Although the subtraction is not as clean in this case due to the difference in instruments, the photometry of the residuals is consistent with the GTC results, indicating an excess in redder optical bands with respect to the afterglow.

We compare the luminosity of the SN of GRB\,100418A with SN\,1998bw by applying a multiplying factor to the supernova template and combining it with the afterglow emission to match the subtracted photometry. In this way, we find that in $i'$-band the SN is 29\% and in $r'$-band 11\% of the luminosity of SN\,1998bw. By subtracting the known contribution of the afterglow to our photometry, we can estimate the magnitude of the supernova associated to GRB\,100418A, at the time of the GTC observation, to be $g'>25.82$ mag, $r'=25.99\pm0.36$ mag and $i'=24.52\pm0.12$ mag. Although the epoch of the GTC observation is slightly after the maximum expected for a SN\,1998bw-like event, the difference between the peak and the observed epoch (for the SN1998bw template) would have been of 0.09 mag in the $r'$-band and 0.04 in the $i'$-band, meaning that these observations should correspond, within errors, to the peak magnitude of the supernova. They are, in any case, the best approximation that we can have for this event. 

At a redshift of $z=0.6239$, the distance modulus of GRB\,100418A corresponds to $\mu=42.91$ mag. With this, we can calculate the absolute magnitudes corresponding to the observed bands to be, in rest frame wavelengths $M_{2900\AA{}}>-17.09$ mag, $M_{3800\AA{}}=-16.92\pm0.36$ mag, $M_{4600\AA{}}=-18.40\pm0.12$ mag. The observer frame $r'$ and $i'$ bands are roughly consistent with rest frame $u'$ and $g'$ bands, respectively.

The limited cadence of the optical light curves at this time limits how much we can say about the observational properties of the associated supernova, i.e. we cannot find the relative stretch and luminosity factors relative to the archetype GRB-SN\,1998bw (e.g. \citealt{zeh04}; \citealt{can14sa}).  
However, we can compare the absolute magnitude of the GRB\,100418A SN with the literature by assuming that the absolute magnitude derived from the $i'$-band observation can be assimilated to a rest frame $V$-band, that once corrected for the line-of-sight extinction results in $M_V=-18.49\pm0.13$ mag.
For example, \citet{ric09} found, for a sample of 14 GRB-SNe, an average peak $V$-band magnitude of $\bar{M_V} = -19.0$~mag with a standard deviation of 0.8~mag. Moreover, referring to the sample of GRB-SNe by \citet{li14}, the faintest GRB-SNe to date were SN\,2010bh ($M_V=-18.89$ mag) and SN~2006aj ($M_V=-18.85$ mag).  As such, the SN associated with GRB~100418A is amongst the faintest, if not the faintest, GRB-SN yet detected.

\begin{figure}
   \centering
   \includegraphics[width=\columnwidth]{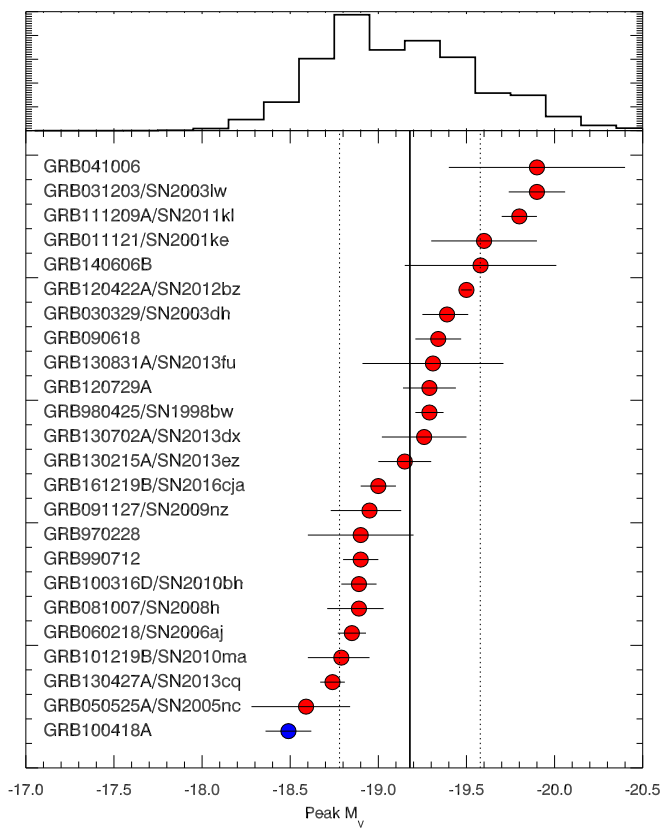}
      \caption{Peak absolute magnitudes in the rest frame $V$-band of a sample of GRB-SNe. The vertical solid line indicates the average of the sample, whereas the dotted lines indicate the standard deviation from this value. The top panel shows a histogram of the $M_V$ distribution. GRB\,100418A, shown in blue, is the faintest supernova within this sample.}
         \label{fig:sn_mv}
   \end{figure}

Using data from \citet{ric09}, \citet{li14}, \citet{can14}, \citet{can15}, \citet{oli15}, \citet{bec17}, \citet{can17}, \citet{can17r}, \citet{kan18} (and references therein) we have either taken the peak $M_V$ calculated in the papers for the corresponding SNe, or derived our own estimate, approximating from the closest band to the rest frame $V$-band. Fig.~\ref{fig:sn_mv} shows a comparison of the magnitudes
of this sample. We obtain that the average value for the sample is $\bar{M_V} = -19.2$~mag, with a standard deviation of 0.4~mag. The average value is slightly brighter than the one derived for the sample of \citet{ric09}, and with a tighter standard deviation. We note that we did not use the 3 SNe that were already marked in the sample of \citet{ric09} as highly uncertain. Within this sample, the SN associated with GRB\,100418A would be the faintest detected to date.

\section{Host galaxy analysis}
\label{host}

\begin{figure}
   \centering
   \includegraphics[width=\columnwidth]{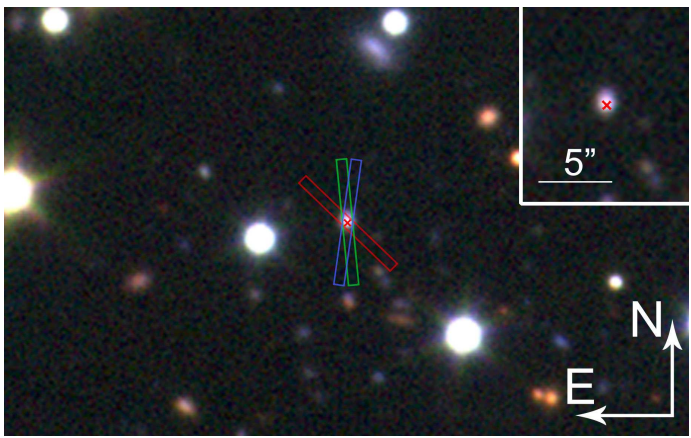}
      \caption{Image of the host galaxy field and the three slit positions of the three different X-shooter epochs (epoch 1 is red, epoch 2 green and epoch 3 is blue). The GRB position is indicated with a cross.}
         \label{Fig:hostpic}
   \end{figure}

The host of GRB\,100418A is a SF dwarf galaxy with a slight elongation in the North-South direction (see Fig.~\ref{Fig:hostpic}). The apparent physical size of the galaxy is $(7.3\pm0.6)\times(6.0\pm0.5)$\,kpc ($1\farcs05\pm0\farcs08\times0\farcs86\pm0\farcs07$ on the sky). Although the galaxy is compact in imaging, spectroscopy reveals that it consists of at least two SF knots aligned not far from the line-of-sight (see Fig. \ref{Fig:Em}) but separated by 130 km s$^{-1}$ in velocity space. The GRB was coincident with the brighter, redshifted, southernmost knot; the fainter elongated feature likely corresponds to the second, blueshifted knot, hence it is in the foreground of the main SF knot. The three X-shooter spectra were taken with slightly different slit angles, hence covering slightly different percentages of the knots. The brightness of both afterglow and emission features allows us to perform a detailed study of the host using both gas emission and interstellar medium (ISM) properties from absorption in the afterglow spectrum.

   \begin{figure}
   \centering
   \includegraphics[width=\columnwidth]{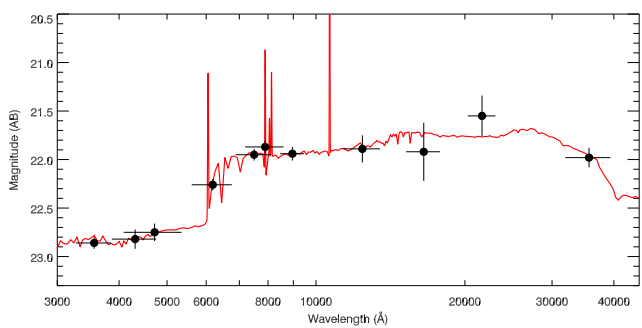}
      \caption{Spectral energy distribution of GRB\,100418A with the best fit from \texttt{LePhare} over-plotted in red.}
         \label{Fig:hostSED}
   \end{figure}
   
 \begin{table}
 \caption{Host galaxy magnitudes, in the AB system, corrected for Galactic extinction, together with the magnitudes derived from the model fit.}
\label{table:hostmags}      %
\begin{center}          
\begin{tabular}{c c c c}  
\hline\hline               
Filter      & Telescope/Instrument & AB$_{obs}$ mag & AB$_{model}$ mag     \\  
\hline  
\hline
$u$         & 10.4m GTC/OSIRIS   & 22.86$\pm$0.06  & 22.829  \\
$B$         & 10m Keck/LRIS      & 22.82$\pm$0.10  & 22.817  \\
$g$         & 10.4m GTC/OSIRIS   & 22.75$\pm$0.09  & 22.775  \\
$r$         & 10.4m GTC/OSIRIS   & 22.26$\pm$0.06  & 22.288  \\
$i$         & 10.4m GTC/OSIRIS   & 21.95$\pm$0.06  & 21.924  \\
$I$         & 10m Keck/LRIS      & 21.87$\pm$0.06  & 21.898  \\
$z$         & 10.4m GTC/OSIRIS   & 21.94$\pm$0.07  & 21.939  \\
$J$         & 3.5m CAHA/O2000    & 21.89$\pm$0.14  & 21.861  \\
$H$         & 3.5m CAHA/O2000    & 21.92$\pm$0.30  & 21.733  \\
$K_S$       & 4.2m WHT/LIRIS     & 21.55$\pm$0.21  & 21.766  \\
$3.6 \mu m$ &0.85m {\it Spitzer}/IRAC&21.98$\pm$0.10& 21.981 \\
\hline\hline
\end{tabular}
\end{center}
\end{table}  
   
\subsection{SED fit of the host galaxy}\label{Sect:SEDfithost}

In Table~\ref{table:hostmags} we list the observed magnitudes from late-time imaging of the host galaxy from {\it u} to {\it 3.6$\mu$m} bands using several different telescopes. We used LePhare \citep[v. 2.2, ][]{arn99,ilb06} to fit the optical-to-NIR SED of the host to a set of galaxy templates based on the models from \citet{bru03}. The SED shape is well-reproduced ($\chi^2/d.o.f.=2.3/8$) by a galaxy template (see Fig.~\ref{Fig:hostSED}) reddened by a Calzetti extinction law with $E(B-V)=0.1$ \citep{cal00}, an age of $203_{-82}^{+334}$ Myr, a mass of log(M$_*$/M$_\odot$)$=9.20_{-0.13}^{+0.10}$, and a star formation rate of log(\textit{SFR}/(M$_\odot$ yr$^{-1})) = 0.8_{-0.2}^{+0.3}$, implying a specific star formation rate of log(\textit{SSFR}/yr$^{-1})= -8.4_{-0.3}^{+0.4}$. The galaxy lies within the average value of GRB host masses at $z\sim0.6$ but has a value for the SFR and hence the SSFR on the upper end of the distribution at that redshift \citep[see e.g.][]{per16,ver15,jap16}. 
 
\subsection{Absorbing gas along the line of sight}\label{Sect:abs}

   \begin{figure*}
   \centering
   \includegraphics[width=13cm]{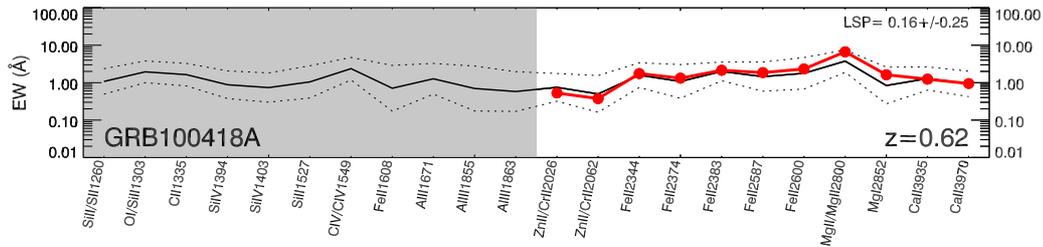}
      \caption{Line strength diagram of the afterglow of GRB\,100418A, following the prescription of \citet{deu12b}. This figure compares the strength of the spectral features of GRB\,100418A (in red) with the average values of a sample of 69 spectra (in black).}
         \label{Fig:LSDiagram}
   \end{figure*}

The X-shooter spectra show absorption features commonly seen in GRB spectra, including \ion{Zn}{II}, \ion{Cr}{II}, \ion{Fe}{II}, \ion{Mn}{II}, \ion{Mg}{II}, \ion{Mg}{I}, \ion{Ti}{II}, \ion{Ca}{II}. In the first spectrum we also detect the \ion{Fe}{II*}$\lambda$2396.36 fine-structure line (see Table~\ref{table:ewlines}), which is thought to be excited by the GRB emission \citep{vre07}. This line, detected with an equivalent width of $0.13\pm0.2$ {\AA} in the first epoch disappears at the time of the second spectrum, where we determine a 3-$\sigma$ detection limit of 0.09 {\AA}. This is consistent with the explanation of having been excited by the GRB, and allows us to unambiguously associate the GRB to the $z=0.6239$ redshift. In Table~\ref{table:ewlines} we measured the combined absorption of each of the observed features in the first epoch, by calculating their equivalent widths. The line-strength diagram (following the prescription of \citealt{deu12b}) in Fig.~\ref{Fig:LSDiagram} shows that the absorption features are almost exactly on the average found for GRB afterglows, with the only exception being magnesium lines, which are slightly higher than average of the sample of \citet{deu12b}. We determine a line-strength parameter of $\textrm{LSP}=0.16\pm0.25$, which implies that the lines are consistent with the average of the sample (consistent with average value of LSP = 0) with a formal value that implies that they are stronger than those of 60\% of the sample. We do not find a significant variation of the line strengths in the different epochs ($\textrm{LSP}=0.04\pm0.32$ in the second epoch and $\textrm{LSP}=0.13\pm0.34$ in the third).

\begin{table}[th]
\caption{List of features identified in the spectra (detections $\gtrsim3\sigma$) along with their observer-frame equivalent widths.}
\label{table:ewlines}      
\centering                          
\begin{tabular}{c c c c}        
\hline\hline                 
\textbf{$\lambda_{\mathbf{obs}}$} & \textbf{Feature} & \textbf{$z$} & \textbf{EW$_{\mathbf{obs}}$}            \\    
{\scriptsize \textbf{(\AA)}} & {\scriptsize \textbf{(\AA)} }  &                          & {\scriptsize \textbf{(\AA)} }   \\  \hline\hline  
3290.12	& \ion{Zn}{II}$\lambda$2026.14   & 0.6238	&  0.67$\pm$0.08\\	
3338.59	& \ion{Cr}{II}$\lambda$2056.26   & 0.6236	&  0.26$\pm$0.05\\	
3348.78	& \ion{Zn}{II}$\lambda$2062.66   & 0.6235	&  0.51$\pm$0.06\\	
3354.91	& \ion{Cr}{II}$\lambda$2066.16   & 0.6237	&  0.19$\pm$0.04\\	
3652.81	& \ion{Fe}{II}$\lambda$2249.88   & 0.6236	&  0.40$\pm$0.05\\	
3670.69	& \ion{Fe}{II}$\lambda$2260.78   & 0.6236	&  0.38$\pm$0.04\\	
3805.58	& \ion{Fe}{II}$\lambda$2344.21   & 0.6234	&  2.81$\pm$0.05\\	
3854.97	& \ion{Fe}{II}$\lambda$2374.46   & 0.6235	&  2.08$\pm$0.04\\	
3868.05	& \ion{Fe}{II}$\lambda$2382.77   & 0.6233	&  3.50$\pm$0.04\\	
3891.45	& \ion{Fe}{II*}$\lambda$2396.36   & 0.6239 	&  0.13$\pm$0.02\\	
3934.57	& \ion{Ca}{II}$\lambda$3934.78   & -0.0001	&  0.39$\pm$0.02\\	
3969.04	& \ion{Ca}{II}$\lambda$3969.59   & -0.0001	&  0.21$\pm$0.02\\	
4183.91	& \ion{Mn}{II}$\lambda$2576.88   & 0.6236	&  0.87$\pm$0.03\\	
4199.24	& \ion{Fe}{II}$\lambda$2586.65   & 0.6234	&  3.02$\pm$0.03\\	
4212.67	& \ion{Mn}{II}$\lambda$2594.50   & 0.6237	&  0.66$\pm$0.02\\	
4221.02	& \ion{Fe}{II}$\lambda$2600.17   & 0.6234	&  3.68$\pm$0.03\\	
4231.96	& \ion{Mn}{II}$\lambda$2606.46   & 0.6236	&  0.59$\pm$0.03\\	
4538.95	& \ion{Mg}{II}$\lambda$2796.35   & 0.6232	&  5.53$\pm$0.04\\	
4550.85	& \ion{Mg}{II}$\lambda$2803.53   & 0.6233	&  5.23$\pm$0.04\\	
4631.90	& \ion{Mg}{I}$\lambda$2852.96   & 0.6235	&  2.76$\pm$0.03\\	
4990.98	& \ion{Ti}{II}$\lambda$3073.88   & 0.6237   &  0.12$\pm$0.02\\	
5265.43	& \ion{Ti}{II}$\lambda$3242.93   & 0.6237	&  0.17$\pm$0.02\\	
5495.43	& \ion{Ti}{II}$\lambda$3384.74   & 0.6236	&  0.15$\pm$0.02\\	
5890.90	& \ion{Na}{I}$\lambda$5891.58    & -0.0001	&  0.41$\pm$0.05\\	
5897.12	& \ion{Na}{I}$\lambda$5897.56    & -0.0001	&  0.23$\pm$0.04\\	
6389.74	& \ion{Ca}{II}$\lambda$3934.78   & 0.6239	&  1.85$\pm$0.04\\	
6446.29	& \ion{Ca}{II}$\lambda$3969.59   & 0.6239	&  1.22$\pm$0.02\\	
\hline\hline
\end{tabular}
\end{table}

   \begin{figure}[]
   \centering
   \includegraphics[width=7.5cm]{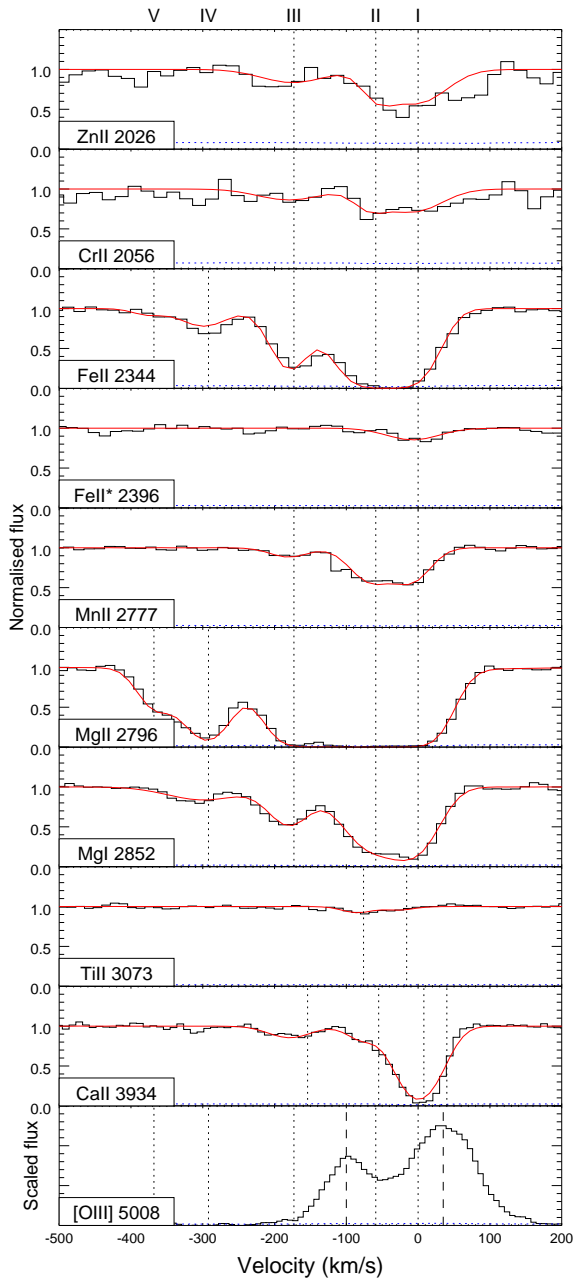}
      \caption{Voigt profile fitting of several absorption features, and the \ion{[O}{III]}$\lambda5008$ emission in the lower panel. The five main absorption components are indicated with roman numbers at the top. In the bottom panel we have also indicated the two main emission components with dashed lines.}
         \label{Figlfit}
   \end{figure}

\begin{table*}
\caption{List of features identified in the first X-shooter spectrum (detections $\gtrsim3\sigma$).}             
\label{table:lines}      
\centering                          
\begin{scriptsize}
\begin{tabular}{c c c c c c c c c c c c}        
\hline\hline                 
 &  & \multicolumn{2}{c}{\textbf{Component I}}& \multicolumn{2}{c}{\textbf{Component II}} & \multicolumn{2}{c}{\textbf{Component III}} & \multicolumn{2}{c}{\textbf{Component IV}} & \multicolumn{2}{c}{\textbf{Component V}} \\    
\hline
    \textbf{Ion}   &\textbf{Transitions}  & \textbf{vel.} &\textbf{log N }  & \textbf{vel.}&\textbf{log N }  &\textbf{vel.}&\textbf{log N }  &\textbf{vel.}&\textbf{log N }  &\textbf{vel.}&\textbf{log N}  \\ 
       &\scriptsize \textbf{ (\AA)}		&\scriptsize \textbf{$\mathbf{km~s^{-1}}$}&\scriptsize \textbf{$\mathbf{cm^{-2}}$}&\scriptsize \textbf{$\mathbf{km~s^{-1}}$}&\scriptsize \textbf{$\mathbf{cm^{-2}}$}&\scriptsize \textbf{$\mathbf{km~s^{-1}}$}&\scriptsize \textbf{$\mathbf{cm^{-2}}$}&\scriptsize \textbf{$\mathbf{km~s^{-1}}$}&\scriptsize \textbf{$\mathbf{cm^{-2}}$} &\scriptsize \textbf{$\mathbf{km~s^{-1}}$}&\scriptsize \textbf{$\mathbf{cm^{-2}}$}\\
\hline\hline  
 \ion{Zn}{II} 	& \scriptsize 2026, 2062  				& +0 &13.31$\pm$0.06	& --59 	& 13.05$\pm$0.21	& --173		&12.87$\pm$0.12 	& ---		& ---		& --- & ---		\\ %
 \ion{Cr}{II} 	& \scriptsize 2056, 2066  				& +0 & 13.75$\pm$0.07   	& --59 	& 13.53$\pm$0.14 	& --173 		& 13.44$\pm$0.10				& --- 		& ---		& --- & ---			\\ %
 \ion{Fe}{II} 	& \scriptsize 2249, 2260, 2344, 2374,	& +0 & 15.19$\pm$0.06   	& --59 	& 15.35$\pm$0.05 	& --173	& 14.49$\pm$0.03	& --292	& 13.46$\pm$0.01 	& --368 & 12.96$\pm$0.06	\\ %
 			& \scriptsize 2382, 2586, 2600			&	&				&		&				&		&				&		&				&  & 	\\
 \ion{Fe}{II*} 	&\scriptsize 2396   					& +0 & 12.86$\pm$0.05   	& --- 		& --- 				& --- 		& ---  			& --- 		& ---		& --- & ---			\\ %
 \ion{Mn}{II} 	& \scriptsize 2576, 2594, 2606  		& +0 & 13.17$\pm$0.02 	& --59 	& 13.22$\pm$0.02 	& --173	& 12.54$\pm$0.07  	& --- 		& ---		& --- & ---			\\ %
 \ion{Mg}{II} 	& \scriptsize 2796, 2803  				& +0 & 16.00$\pm$0.17   	& --59 	& 16.61$\pm$0.17 	& --173	& 14.71$\pm$0.15 	& --292 	& 13.78$\pm$0.08 	& --368 & 13.06$\pm$0.03	\\ %
 \ion{Mg}{I} 	& \scriptsize 2852  					& +0 & 13.10$\pm$0.02   	& --59	& 12.98$\pm$0.01 	& --173	& 12.58$\pm$0.01 	& --292 	& 12.20$\pm$0.0.3	& --- & --- 	\\ %
 \ion{Ti}{II} 	& \scriptsize 3073, 3242, 3384  		& --16 & 12.76$\pm$0.47  			& --76 	& 12.73$\pm$0.29 	& --- 		& ---  			& --- 		& ---		& --- & ---			\\ %
 \ion{Ca}{II} 	& \scriptsize 3934, 3969  				& +40 & 13.38$\pm$0.03 	& +8	& 15.19$\pm$0.22 	& --55	& 12.89$\pm$1.00	& --154 		& 12.21$\pm$0.03		& --- & ---			\\ %
\hline\hline
\end{tabular}
\end{scriptsize}
\end{table*}

Using the first spectrum with the highest S/N for the absorption lines we perform a Voigt-profile fitting of the absorption features, using \texttt{FITLYMAN} \citep{fon95} in the MIDAS environment. The results of the fits are shown in Table~\ref{table:lines}. In the strongest features we identify up to five different velocity components at $\textrm{v}=0$ (component I), $-59,\,-173,\,-292$ and $-368$ km\,s$^{-1}$ (component V). Only the strongest absorption lines such as \ion{Fe}{II} and \ion{Mg}{II} + \ion{Mg}{I} have absorption in all components (component V cannot be clearly identified in \ion{Mg}{I}). \ion{Mn}{II}, \ion{Zn}{II} and \ion{Cr}{II} are only present in components I and II, corresponding to the brighter SF region of the galaxy, although the velocities of \ion{Zn}{II} and \ion{Cr}{II} are slightly shifted. \ion{Ca}{II} has a somewhat different velocity structure than the other lines but also shows a weak component within the velocities corresponding to the second SF region. 

The \ion{Fe}{II*} fine-structure line is rather weak and only detected in the component at $\textrm{v}=0$\,km\,s$^{-1}$ and disappears between the first and second epoch. This line was probably excited by the emission of the GRB via UV pumping or photo-excitation \citep[see e.g.][]{vre07,del09,vre13}. It also marks the closest velocity component to the GRB and hence determines by definition the redshift of the GRB and relative velocity $\textrm{v}=0$\,km\,s$^{-1}$. We checked the different epochs for any variation in the strength of the resonant absorption lines but do not find any evidence. The absorbing column density of the disappearing \ion{Fe}{II*} fine-structure line would show up as additional absorption in the corresponding resonant transition.  However, the absorption in \ion{Fe}{II} in this component is two dex higher than the \ion{Fe}{II*} column density, is blended with component II, and even in the weakest transition of \ion{Fe}{II} ($\lambda$ 2260 \AA{}) mildly saturated. Hence making the uncertainty in the fit larger than any additional absorption in component I of \ion{Fe}{II} (adding the absorbing column of \ion{Fe}{II*} onto component I would add $<0.01$ dex to the column density of \ion{Fe}{II}). The data of the third epoch have too low S/N for any further constraints due to the decrease of the afterglow continuum.

\subsection{Host galaxy emission lines}\label{sec:emlines}

  \begin{figure}
   \centering
   \includegraphics[width=\columnwidth]{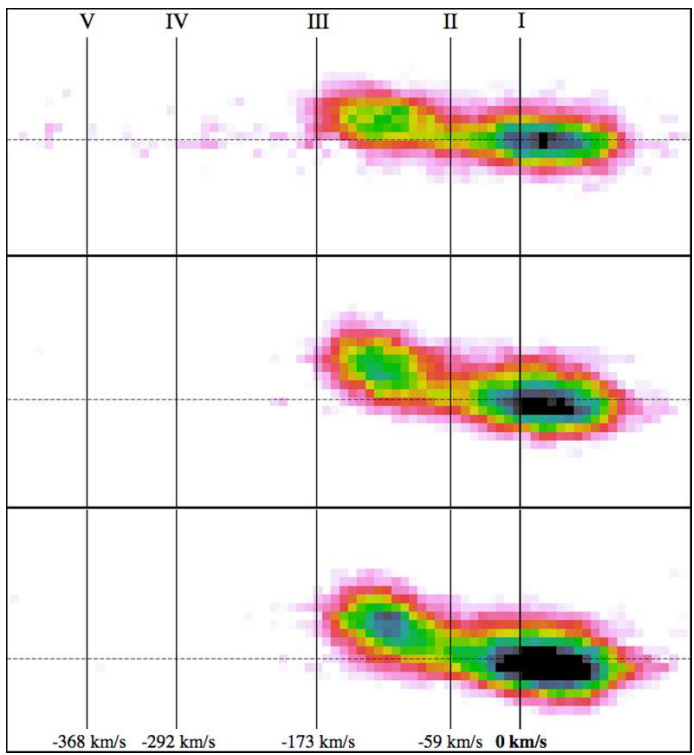}
      \caption{Section of the 2D spectra showing the \ion{[O}{III]}$\lambda$5008.24 emission line, with continuum subtracted, for each of the three spectral epochs (from top to bottom). The spatial axis is vertical, while the spectral dispersion is horizontal. The northernmost part of the slit is to the top (see Fig.~\ref{Fig:hostpic} for each slit orientation). The dashed line marks the position of the GRB afterglow in the spatial direction. Horizontal lines indicate the relative velocities of the 5 main absorption components, some of which are related to the emission features.}
         \label{Fig:Em}
   \end{figure}

Already in the first spectrum, still dominated by the afterglow, the emission lines from the host galaxy are well-detected. However, in the following analysis we use the spectrum of the third epoch at 2.5 d post-GRB, unless otherwise mentioned, which has higher S/N in the emission lines due to a lower afterglow contribution and a better alignment of the slit comprising most of the flux of the host galaxy (see Fig. \ref{Fig:Em}). We detect at least 17 emission lines as detailed in Table~\ref{table:emlines}, including Balmer and Paschen lines, \ion{[O}{II]}, \ion{[O}{III]}, \ion{[Ne}{II]}, \ion{[N}{II]}, \ion{[S}{II]}, \ion{[S}{III]}, and \ion{[He}{I]}.  These features all show two main velocity components which, for the higher S/N lines, can be separated into further subcomponents, which we will detail in Sect. \ref{sect:kinematics}
Table~\ref{table:emlines} displays the measurement of the emission lines observed in the spectrum. For each of the two components, we show the full-width half-maximum, central wavelength and flux resulting from a double Gaussian deblend procedure. The last two columns show the central wavelength and flux of the combined lines, measured without assuming any line profile.

\begin{table*}
\caption{List of emission lines and their fluxes measured in the different spectra.
Fluxes are corrected for Galactic extinction and normalised to the photometry. 
No intrinsic extinction corrections have been applied.}
\label{table:emlines}      
\centering                          
\begin{scriptsize}
\begin{tabular}{c c c c c}        
\hline\hline                 
Line	& Flux$_\mathbf{1}$	& Flux$_{2}$	& Flux$_{1+2}$ & Flux$_{Comb}$ 	\\
{\AA}           & ($10^{-17}$ erg s$^{-1}$ cm$^{-2}$)& ($10^{-17}$ erg s$^{-1}$ cm$^{-2}$)& ($10^{-17}$ erg s$^{-1}$ cm$^{-2}$)& ($10^{-17}$ erg s$^{-1}$ cm$^{-2}$)\\
\hline\hline
                                &                   & EPOCH 1           &               &   \\
\hline\hline
 \ion{[O}{II]}$\lambda$3727.09  & 6.04$\pm$0.55     & 9.61$\pm$0.79     & 15.65$\pm$0.96    & 41.89$\pm$0.95    \\	
\ion{[O}{II]}$\lambda$3729.87	& 9.11$\pm$0.64     & 14.78$\pm$0.80    & 23.89$\pm$1.02    & blend		        \\
\ion{[Ne}{II]}$\lambda$3870.87	& ---		        & ---		        & ---	            & ---	            \\
\ion{H}{6}$\lambda$3890.15		& 0.23$\pm$0.50     & 1.78$\pm$0.63     & 2.01$\pm$0.80     & 2.48$\pm$0.81	    \\
\ion{H}{5}$\lambda$3971.20		& ---		        & ---		        & ---               & ---	            \\
\ion{H}{4}$\lambda$4102.89		& 0.85$\pm$0.43     & 1.90$\pm$0.54     & 2.74$\pm$0.69     & 2.96$\pm$0.60	    \\	
\ion{H}{3}$\lambda$4341.68		& 1.77$\pm$0.41     & 3.37$\pm$0.52     & 5.14$\pm$0.66     & 5.16$\pm$0.93	    \\
\ion{H}{$\beta$}$\lambda$4862.68& 4.46$\pm$0.46     & 9.21$\pm$0.61     & 13.67$\pm$0.76    & 14.30$\pm$0.70    \\
\ion{[O}{III]}$\lambda$4960.30	& 3.35$\pm$0.42     & 7.46$\pm$0.50     & 10.81$\pm$0.65    & 11.09$\pm$0.72    \\
\ion{[O}{III]}$\lambda$5008.24	& 10.09$\pm$0.40    & 20.21$\pm$0.62    & 31.12$\pm$0.62    & 32.76$\pm$0.69    \\
\hline		
\ion{H}{$\alpha$}$\lambda$6564.61& 14.40$\pm$1.43   & 25.07$\pm$2.02    & 39.47$\pm$2.48	& 41.19$\pm$1.25    \\
\ion{[N}{II]}$\lambda$6585.28	& 1.22$\pm$1.58		& 2.35$\pm$2.44	    & 3.57$\pm$2.91		& (2.76$\pm$1.5)    \\
\ion{[S}{II]}$\lambda$6718.29	& 1.34$\pm$1.47		& 3.45$\pm$1.57	    & 4.80$\pm$2.16		& 5.01$\pm$1.33     \\
\ion{[S}{II]}$\lambda$6732.67	& 0.94$\pm$1.34		& 1.34$\pm$1.71	    & 2.27$\pm$2.16		& 5.77$\pm$1.26     \\
\ion{[S}{III]}$\lambda$9533.20	& 2.52$\pm$0.82		& 4.88$\pm$0.72	    & 7.40$\pm$1.08		& 6.57$\pm$1.28     \\
\ion{He}{I}$\lambda$10833.3		& ---				& ---			    & ---			    & ---               \\
\ion{Pa}{$\gamma$}$\lambda$10941.10 & ---			& ---			    & ---			    & ---               \\
\hline\hline
                                &                   & EPOCH 2           &                   &                   \\
\hline\hline
\ion{[O}{II]}$\lambda$3727.09	& 6.94$\pm$0.52		& 14.28$\pm$0.78	& 21.22$\pm$0.94	& 53.32$\pm1.24$    \\	
\ion{[O}{II]}$\lambda$3729.87   & 9.87$\pm$0.55		& 19.03$\pm$0.79	& 28.9$\pm$0.96		& blend		        \\
\ion{[Ne}{II]}$\lambda$3870.87  & 1.31$\pm$0.41		& 2.06$\pm$0.64		& 3.37$\pm$0.76		& 3.58$\pm$0.91     \\
\ion{H}{6}$\lambda$3890.15		& 1.27$\pm$0.41		& 1.54$\pm$0.58		& 2.81$\pm$0.71		& 3.24$\pm$0.41     \\
\ion{H}{5}$\lambda$3971.20		& --- (sky line)	& 2.09$\pm$0.69		& $>2.09\pm$0.81	& 2.16$\pm$0.33     \\
\ion{H}{4}$\lambda$4102.89		& 1.13$\pm$0.40		& 3.09$\pm$0.56		& 4.23$\pm$0.69		& 4.82$\pm$0.57     \\
\ion{H}{3}$\lambda$4341.68		& 1.96$\pm$0.35		& 4.68$\pm$0.49		& 6.63$\pm$0.60		& 7.20$\pm$0.30     \\
\ion{H}{$\beta$}$\lambda$4862.68& 5.10$\pm$0.41		& 11.58$\pm$0.54	& 16.68$\pm$0.68	& 17.35$\pm$0.31     \\
\ion{[O}{III]}$\lambda$4960.30  & 4.18$\pm$0.34		& 9.19$\pm$0.48		& 13.37$\pm$0.59 	& 14.99$\pm$0.45     \\
\ion{[O}{III]}$\lambda$5008.24  & 13.36$\pm$0.33	& 28.27$\pm$0.53	& 41.62$\pm$0.62	& 41.79$\pm$1.06     \\
\hline		     
\ion{H}{$\alpha$}$\lambda$6564.61& 14.75$\pm$1.65	& 31.37$\pm$1.65	& 46.12$\pm$2.87	& 52.64$\pm$1.96     \\
\ion{[N}{II]}$\lambda$6585.28	& 0.69$\pm$1.86		& 1.64$\pm$2.95		& 2.33$\pm$3.49		& 4.23$\pm$1.02     \\
\ion{[S}{II]}$\lambda$6718.29	& 3.08$\pm$1.68		& 5.00$\pm$1.80		& 8.07$\pm$2.46		& 10.75$\pm$2.35     \\
\ion{[S}{II]}$\lambda$6732.67	& 1.50$\pm$1.54		& 2.18$\pm$1.99		& 3.68$\pm$2.52		& 6.41$\pm$1.05     \\
\ion{[S}{III]}$\lambda$9533.20	& 3.13$\pm$0.75		& 5.83$\pm$0.70		& 8.97$\pm$1.03		& 8.64$\pm$1.23     \\
\ion{He}{I}$\lambda$10833.3		& 1.19$\pm$0.55		& 1.03$\pm$0.92		& 2.23$\pm$1.07		& 1.69$\pm$1.00     \\
\ion{Pa}{$\gamma$}$\lambda$10941.10& 	---			& ---				& ---	            & ---               \\
\hline\hline
                                &                   & EPOCH 3           &               &   \\
\hline\hline
\ion{[O}{II]}$\lambda$3727.09	& 7.26$\pm$0.43	    & 12.96$\pm$0.58	& 20.22$\pm$0.72	& 49.74$\pm$0.70     \\	
\ion{[O}{II]}$\lambda$3729.87	& 10.21$\pm$0.52	& 17.92$\pm$0.66	& 28.13$\pm$0.84	& blend		         \\
\ion{[Ne}{II]}$\lambda$3870.87	& 1.23$\pm$0.33		& 2.17$\pm$0.47		& 3.41$\pm$0.57		& 3.28$\pm$0.42     \\
\ion{H}{6}$\lambda$3890.15		& 0.24$\pm$0.37		& 1.53$\pm$0.47		& 1.77$\pm$0.60		& 1.94$\pm$0.37     \\
\ion{H}{5}$\lambda$3971.20		& 0.68$\pm$0.39		& 1.22$\pm$0.50		& 1.90$\pm$0.64		& 2.02$\pm$0.37     \\
\ion{H}{4}$\lambda$4102.89		& 0.84$\pm$0.32		& 2.32$\pm$0.46		& 3.16$\pm$0.56		& 3.11$\pm$0.48     \\
\ion{H}{3}$\lambda$4341.68		& 1.87$\pm$0.28		& 4.35$\pm$0.41		& 6.22$\pm$0.50		& 6.10$\pm$0.72     \\
\ion{H}{$\beta$}$\lambda$4862.68& 5.22$\pm$0.37		& 10.79$\pm$0.49	& 16.01$\pm$0.61 	& 16.37$\pm$0.46     \\
\ion{[O}{III]}$\lambda$4960.30	& 3.99$\pm$0.32		& 8.88$\pm$0.38		& 12.87$\pm$0.50	& 13.66$\pm$0.32     \\
\ion{[O}{III]}$\lambda$5008.24	& 12.51$\pm$0.30	& 26.39$\pm$0.44	& 38.90$\pm$0.53	& 39.25$\pm$0.39     \\
\hline		
\ion{H}{$\alpha$}$\lambda$6564.61& 19.75$\pm$1.65	& 34.06$\pm$1.94	& 53.82$\pm$2.55	& 56.43$\pm$1.00     \\
\ion{[N}{II]}$\lambda$6585.28	& 3.08$\pm$1.93		& 5.91$\pm$2.42		& 8.98$\pm$3.10		& 6.03$\pm$1.19     \\
\ion{[S}{II]}$\lambda$6718.29	& 1.99$\pm$1.28		& 4.53$\pm$1.47		& 6.52$\pm$1.95		& 7.70$\pm$1.01     \\
\ion{[S}{II]}$\lambda$6732.67	& 2.64$\pm$1.53		& 3.74$\pm$1.67		& 6.38$\pm$2.27		& 6.75$\pm$1.41     \\
\ion{[S}{III]}$\lambda$9533.20	& 2.26$\pm$0.80		& 5.34$\pm$0.68		& 7.60$\pm$1.05		& 8.33$\pm$1.01     \\
\ion{He}{I}$\lambda$10833.3		& 1.49$\pm$0.52		& 2.10$\pm$0.93		& 3.59$\pm$1.07		& 4.11$\pm$1.02     \\
\ion{Pa}{$\gamma$}$\lambda$10941.10& 0.47$\pm$0.87	& 1.26$\pm$0.97		& 1.73$\pm$1.30		& 2.77$\pm$1.00     \\
\hline\hline
\end{tabular}
\end{scriptsize}
\end{table*}

Apart from the \ion{He}{I} $\lambda$ 10833 line we do not detect any other transitions of \ion{He}{I} such as the lines at $\lambda_\mathrm{rest}$=3819, 4026, 4471, 4922, 5019, 5876, 6678 or 7065 \AA{} nor ionised \ion{He}{II} at $\lambda$ 3204, 4686 or 10126 \AA{}. The \ion{He}{I} $\lambda$ 10833 emission line is the transition from the He-triplet state 2 $^3$P$_0$ (1s2p) to 2 $^3$S (1s2s) and can be populated either by collisions from the 2 $^3$S or after recombination in the downward cascade from a higher excited state e.g. via emission of $\lambda\lambda$ 4471, 5876 or 7065 \AA{} photons, none of which we detect \citep[for a Grotrian diagram see e.g.][]{ben99}. Due to the additional population by collisions, the ratio of e.g. \ion{He}{I} 10833 to 4471 increases at higher densities and temperatures and can vary between a few and more than 40 K \citep{ben99}. We determine flux limits for \ion{He}{I} 4771 and 7065 (the 5876 line is in the middle of a skyline) of $<1.01\times10^{-17}$ and $<1.06\times10^{-16}$ erg s$^{-1}$ cm$^{-2}$, respectively. The tighter limit for \ion{He}{I} 4771 together with the flux measured for combined \ion{He}{I} 10833 (see Table\ref{table:emlines}) gives limits on temperatures and densities less than 5000 K and $n_e=10^2\;\textrm{cm}^{-2}$ according to the values tabulated in \cite{ben99}. 

\ion{He}{I} is only visible in emission a few Myr after the onset of the starburst which allows for a rather accurate age determination of the stellar population \citep{gon99}. \ion{He}{I} has been detected in a few nearby GRB hosts (GRB 060218, \citealt{wie07}; GRB 031203, \citealt{mar07, gus11}; GRB 100316D, \citealt{izz17}; and possibly GRB 120422A, \citealt{sch14}) but only in the VIS regime due to lack of high S/N infrared spectra with the exception of the X-shooter spectrum of GRB 031203 \citep{gus11}. However, the dependence of the $\lambda$ 10833 transition on temperature and density (due to population by collisions) makes it very little suited as an age indicator for the stellar population in contrast to the VIS transitions of \ion{He}{I}. \ion{He}{II}, which we do not detect, requires not only a young population but also a strongly ionising radiation field and has been observed to be stronger at low metallicities \citep{sch03}. Some have associated the emission of narrow (nebular) He II emission to Wolf-Rayet (WR) stars but they are not always spatially coincident and WR stars \citep{keh15} are much rarer at low metallicities. For \ion{He}{II} 3204 and 10126 we derive limits of $<1.53\times10^{-17}$ and $<9.3\times10^{-17}$ erg s$^{-1}$ cm$^{-2}$, respectively, \ion{He}{II} $\lambda$ 4686, unfortunately, is in the atmospheric A band, so no useful limits can be derived.  A few other GRB hosts show \ion{He}{II} emission \citep[see e.g.][]{han10} but at luminosities lower than inferred from our detection limit. As the afterglow was still bright during the last X-shooter epoch, the non-detection is not surprising nor constraining.

The electron density is commonly derived using the ratio of the \ion{[S}{II}] $\lambda\lambda$ 6716, 6732 or the \ion{[O}{II}]$\lambda\lambda$3727, 3729 doublets. Since the \ion{[O}{II}] is partially blended due to the extended components, we derive electron temperatures from the \ion{[S}{II}] doublet using a recent recalibration of \cite{ost06} by \cite{pro14}. At $\textrm{T}_e=10,000\;\textrm{K}$, the ratios of I$_{6716}$/I$_{6732}$ result in electron densities between $n_e\sim$ 10$^3$--10$^2$ cm$^{-2}$ for the components and the entire host. For other temperatures, the electron density only changes by a small value. The result is consistent with the limit derived from the ratio of \ion{He}{I} 10830/4471 (see above).

\begin{figure}
   \centering
   \includegraphics[width=\columnwidth]{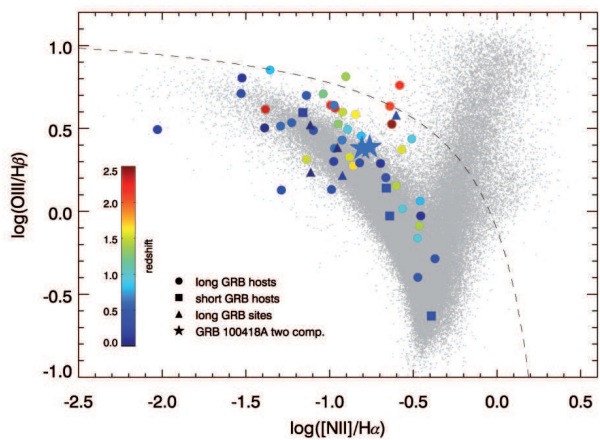}
      \caption{BPT diagram for long and short GRB hosts and GRB sites from the literature compared to SDSS galaxies (grey dots) and the two SF regions in the host of GRB 100418A (stars). The redshift of the hosts is colour-coded in the plotted symbols. Emission line fluxes used are taken from \citet{chr08,ber09,han10,lev10,per12,sch14,tho14,kru15,sta15,ver17,izz17}.}
         \label{fig:bpt}
   \end{figure}

\ion{[S}{III}] $\lambda\lambda$ 9069, 9532 can be used as indicator for excitation of the ISM and as a tracer to  distinguish between HII regions and AGN activity \citep{diaz85}. We do not detect the $\lambda$ 9069 line but the \ion{[S}{III}] doublet has a fixed ratio of 2.486 \citep{all84}.  The host of GRB 100418A lies on the edge of the region in the plot occupied by HII regions in \cite{diaz85} as the fluxes of \ion{[S}{III}] and \ion{[S}{II}] are comparable in value. For HII regions and starbursts, the flux of the \ion{[S}{III}] $\lambda\lambda$ 9069, 9532 is usually higher than the flux of the \ion{[S}{II}] doublet. This could either mean a low ionisation or an enhanced \ion{[S}{II}] flux, indicative of shocked regions or AGN activity, which seems unlikely when placing the galaxy in the BPT diagram (see Fig. \ref{fig:bpt}). We can also derive a sulphur abundance from S$_{23}$=(\ion{[S}{III}]\,$\lambda\lambda$ 9069, 9532\,+\,\ion{[S}{II}]\,$\lambda\lambda$ 6017, 6732)/H$\beta$ adopting the calibration 12 + log(S/H)$=$6.540\,+\,2.071~log S$_{23}$\,+\,0.348~log S$_{23}^2$ from \cite{pm06} and obtain 12+log(S/O)$\sim$ 7.3, a relatively high value but consistent with the high oxygen abundance in the host of GRB 100418A as the ratio S/O has been shown not to vary with metallicity, hence S and O are tightly related \citep[see e.g.][]{keh06}.

\ion{[Ne}{III}] is usually tightly related to \ion{[O}{III}] $\lambda$ 5008 as its isoelectronic equivalent, however, \ion{[Ne}{III}] has a somewhat higher ionisation potential and excitation energy. Also, Ne and O are both $\alpha$ elements produced in Type II SNe and hence related to massive star formation. In both components and the total line flux, we measure log(\ion{[Ne}{III}]/\ion{[O}{III}]) = -1.09, -1.15 and -1.13, consistent with values around -1.1 found for local SDSS galaxies and independent of T$_e$, stellar mass or metallicity \citep{izo06, zei15}. The ratio of log(\ion{[Ne}{III}]/\ion{[O}{III}])$_\mathrm{tot}=-1.2$ is also consistent with the values found for nearby galaxies and the stellar mass derived in Sect. \ref{Sect:SEDfithost} of log $\textrm{M}^*=9.2$ M$_\odot$. At redshifts $z>1$, both ratios are enhanced by $\sim0.2$ dex, the reason for which is still debated, possibilities being Ne enhancement from WR stars, higher densities or higher oxygen depletion \citep{zei15}.

\begin{table}
\caption{Properties for the two main host regions and the entire host complex derived from different emission lines in the spectra of the third epoch. For details, we refer to Sect.\ref{sec:emlines} and \ref{sec:sfr}. The afterglow was coincident with Region 2.}
\label{tab:emlineproperties}      
\centering
\small{
\begin{tabular}{l  c c c}        
\hline\hline                 
\textbf{Property}	& \textbf{Region 1}  	& \textbf{Region 2}  	& \textbf{Combined host}\\ \hline\hline
SFR H$\alpha$ (M$_\odot)$/yr)&3.8$\pm$0.3   &8.3$\pm$0.5   &12.2$\pm$0.6   \\\vspace{1mm}
E(B--V) (mag)& 0.24$\pm$0.09 & 0.08$\pm$0.06  & 0.16$\pm$0.03  \\ \vspace{1mm}
12+log(O/H)& $8.53_{-0.13}^{+0.14}$ & $8.63_{-0.05}^{+0.05}$ & $8.545_{-0.015}^{+0.016}$   \\
n$_e$ (cm$^{-2}$)& 10$^3$  & 1.02$\times$10$^2$  & 1.08$\times$10$^2$  \\
S$_\mathrm{23}$&1.75$\pm$0.28 &2.11$\pm$0.16  & 1.82$\pm$0.14  \\
log [NII]/H$\alpha$&--1.07$\pm$0.27   &--1.03$\pm$0.18  &--1.17$\pm$0.15   \\
log [SII]/H$\alpha$&--0.63$\pm$0.20   &--0.61$\pm$0.09   &--0.59$\pm$0.08   \\
log [OIII]/H$\beta$&0.35$\pm$0.03  &0.43$\pm$0.03   &0.36$\pm$0.02   \\
log N/O&--0.37$\pm$0.27   & --0.34$\pm$0.18      &--0.35$\pm$0.15       \\
\hline\hline
\end{tabular}
}
\end{table}

\subsection{Host galaxy extinction and SFR}\label{sec:sfr}

To derive the properties of the host galaxy we analysed the two main emission components, their sum, as well as the combined spectrum. The fluxes of the individual lines were determined by fitting them with Gaussian functions, whereas the measurements for the combined spectrum were obtained by measuring the flux above the continuum without assuming any line shape. Hence, the integrated flux gives us more accurate measurements but does not allow us to deblend the different components within the host. We estimated the interstellar extinction of the host galaxy from the Balmer decrement, adopting the \cite{cal00} attenuation curve with $R_V=4.05$ which is commonly observed in starburst galaxies. We assume a standard recombination model for SF galaxies and the Case B of HI recombination lines \citep{ost06} corresponding to an electron temperature of T=10$^4$K and density of $\textrm{n}_e=10^2$ cm$^{-3}$. This results in an extinction of E(B--V)$_{1, 2, com}=0.24\pm0.09$, $0.08\pm0.06$, $0.16\pm0.03$ mag, for the first, the second component and combined flux including the entire galaxy, respectively. We adopt the extinction value determined from the 3rd epoch for all spectra, since the higher S/N ratio gives us a better accuracy. 

We use the luminosity of H$\alpha$ and the \ion{O}{II}$\lambda$3727,3729 nebular lines as the tracers of SFR by converting the luminosities of the two lines following the \cite{ken94} relations. For this calculation we assume a Salpeter initial mass function (IMF) (N(m)~$\propto$~m$^{-2.35}$) with masses of $0.1-100\,\textrm{M}_{\odot}$, solar metallicity, and continuous star formation. The H$\alpha$ recombination line directly couples the nebular emission with massive star formation, but has some significant limitations: The method is especially sensitive to extinction and IMF uncertainties and works under the assumption that the ionised gas is a tracer of all the massive star formation. For the escape fraction of ionising radiation an upper limit of 3\% has been established \citep{lei95,tan18}. The properties of galaxies derived using a Salpeter Initial Mass Function (IMF) slope are consistent with the stellar population observed in nearby galaxies \citep{mas98} so we expect that the chosen IMF slope does not affect the results significantly. 

In contrast, the forbidden \ion{[O}{II]}$\lambda$3727, 3729 doublet is not directly coupled to the ionising flux but is sensitive to the gas abundance and ionisation state and the excitation of \ion{[O}{II]} can be calibrated empirically through H$\alpha$. The mean ratio of \ion{[O}{II]}$\lambda$3727/H$\alpha$ in an individual galaxy varies between 0.5 and 1.0 dex reflecting the differences between samples used to derive the SFR calibration from \ion{[O}{II]}$\lambda$3727 \citep{gal89,ken92}. Here we adopt the average calibration. The values in both SFRs reported below have been calculated after correcting for extinction. We then obtain a SFR of $12.2\pm0.6$ M$_{\odot}$~year$^{-1}$ from H$\alpha$ and 6.085 M$_{\odot}$~year$^{-1}$ from \ion{[O}{II]} for the entire host galaxy. This discrepancy is probably due to an overcorrection of the extinction in the \ion{[O}{II]} line, which being in the blue is more affected by extinction uncertainty. Because of this, we use the H$\alpha$ value as the reference one. We note that this SFR is consistent with the value determined from the SED fit, but significantly above the one determined by \citet{kru15}, who estimate SFR = 4.2$_{-0.8}^{+1.0}$ M$_{\odot}$~year$^{-1}$. The difference with the values from \citet{kru15}, who used the same data set are due to a different extraction of the spectrum and matched photometry flux calibration correction, since the extinction is equal in both cases.

\subsection{Host galaxy abundances and metallicity}

In order to estimate metallicities (Z) we used the Python code \texttt{pyMCZ} \citep{bia16} that determines oxygen abundance using strong emission-lines standard metallicity diagnostics. The code derives the statistical oxygen abundance confidence region via Monte Carlo simulations. Various emission-line ratios are used in up to 15 theoretical/empirical/combined metallicity calibrations. For the metallicity analysis of the host we used the following strong emission-line fluxes: \ion{[O}{II]}$\lambda$3727/3729, H$\beta$, \ion{[O}{III]}$\lambda\lambda$4959, 5007, H$\alpha$, \ion{[N}{II]}$\lambda$6584, \ion{[S}{II]}$\lambda\lambda$6717, 6731, \ion{[S}{III]}$\lambda$9532. \ion{[O}{III]} $\lambda$ 4363 is not detected in the spectra, likely due to the relatively high metallicity of the system (see below). The values of line fluxes were corrected for both Galactic and intrinsic reddening before calculating the metallicities. The final adopted metallicity for the three different components are listed in Table \ref{tab:emlineproperties}. The full set of metallicity measurements using different calibrators can be found in Table \ref{tabz} in the appendix. For a detailed discussion on the different calibrators and their abbreviations below we refer to the references in the table.

There has been a long-standing debate on which diagnostic should be adopted to have the best metallicity estimate. It is also known that there are systematic offsets between different calibrators \cite[see e.g.][]{kew08}. In order to properly compare the oxygen abundances provided by different methods using different sets of emission lines we used the metallicity conversion determined by \cite{kew08} which allows the different metallicities to be converted into the same base calibration (in this case \cite{kew02}, henceforth KD02). We present the converted metallicities in the bottom part of Table \ref{tabz} for each epoch. All but one method give results within 0.1 dex after converting to the base metallicity. The only exception is the M91 calibration where we use the \ion{[N}{II]}$\lambda$6584/\ion{[O}{II]}$\lambda$3727 line ratio in order to break the R23 degeneracy. This method directly depends on the very weak \ion{[N}{II]}$\lambda$6584 line and indeed, the weaker this line is, independent of the value of the \ion{[N}{II]}$\lambda$6584/OII $\lambda$3727 ratio, the more likely it is to lead to the lower branch of R23 results (with much lower metallicities). 

We finally adopt as values for metallicity those estimated from the 3rd epoch, which has the least afterglow contamination. The mean values after the base metallicity conversion are $12+\log(\textrm{O/H})=8.53^{+0.14}_{-0.13}$, $8.63\pm0.05$ and  $8.55\pm0.02$, for the 1st and the 2nd component and for the entire galaxy (combined flux), respectively. Hence, despite the part of the galaxy hosting the GRB having a slightly lower metallicity, the metallicities of both parts are consistent within errors.

In Fig. \ref{fig:bpt} we plot the position of the two SF regions in a BPT diagram, together with other long and short GRB hosts and GRB sites, colour-coded by redshift. There is no obvious trend with redshift among long GRB hosts in their location in the BPT diagram, indicating that the ionisation conditions in GRB hosts might not be redshift-dependent. The two parts of the host galaxy are located well among other long GRB hosts and sites within the SF branch of the BPT diagram, but not in a very extreme region. As mentioned above, the line ratios clearly indicate excitation by massive stars and not by an AGN. Both parts of the host are at a very similar location, as is their metallicity and other properties, which means that they are two SF regions in the same galaxy and not an interacting system that would usually show larger differences.

\begin{figure}
   \centering
   \includegraphics[width=\columnwidth]{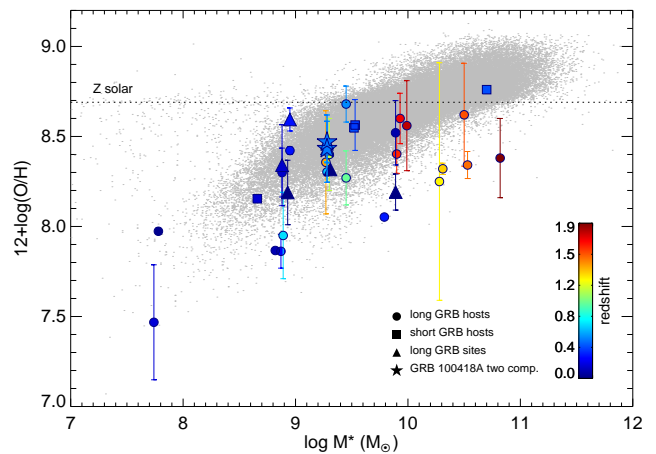}
      \caption{Mass-metallicity relation for GRB hosts and the two SF regions in the host of 100418A. Grey dots are galaxies from the SDSS DR 10. References are the same as in Fig. \ref{fig:bpt} with some additional metallicities not derived via the N2 calibrator taken from \cite{ver17} and \cite{kru15}.}
         \label{fig:mz}
   \end{figure}

In Fig. \ref{fig:mz} we plot the mass-metallicity relation for GRB hosts, including the two SF regions of the host. In contrast to many other long GRB hosts, which usually fall below the mass-metallicity relation, the reason for which has been extensively discussed in many papers \citep[see e.g.][]{ver17}, the host of GRB 100418A is fully consistent with the low-redshift mass-metallicity relation for SDSS galaxies. This is surprising considering that low metallicity - at least at the GRB site - seems to be a common trait among GRB hosts. GRB progenitor models require them to have high angular momentum to produce an accretion disk and launching a jet upon core collapse. However, they need to have retained their fast rotation throughout a phase where they lost their outer layers, which had to have happened given the fact that their accompanying SNe are of Type Ic. The metallicity of the host of 100418A is almost solar and only a few other hosts have had such relatively high metallicities (see Fig. \ref{fig:mz}). Why the host was still able to produce a GRB progenitor we can only speculate about, e.g. the star could have been stripped of its outer layers by a binary companion, a model that has gained popularity in the last years given that most massive stars are in binary systems \citep{san12,lan12}.

\subsection{Host-galaxy kinematics}\label{sect:kinematics}

  \begin{figure}
   \centering
    \includegraphics[width=6.5cm]{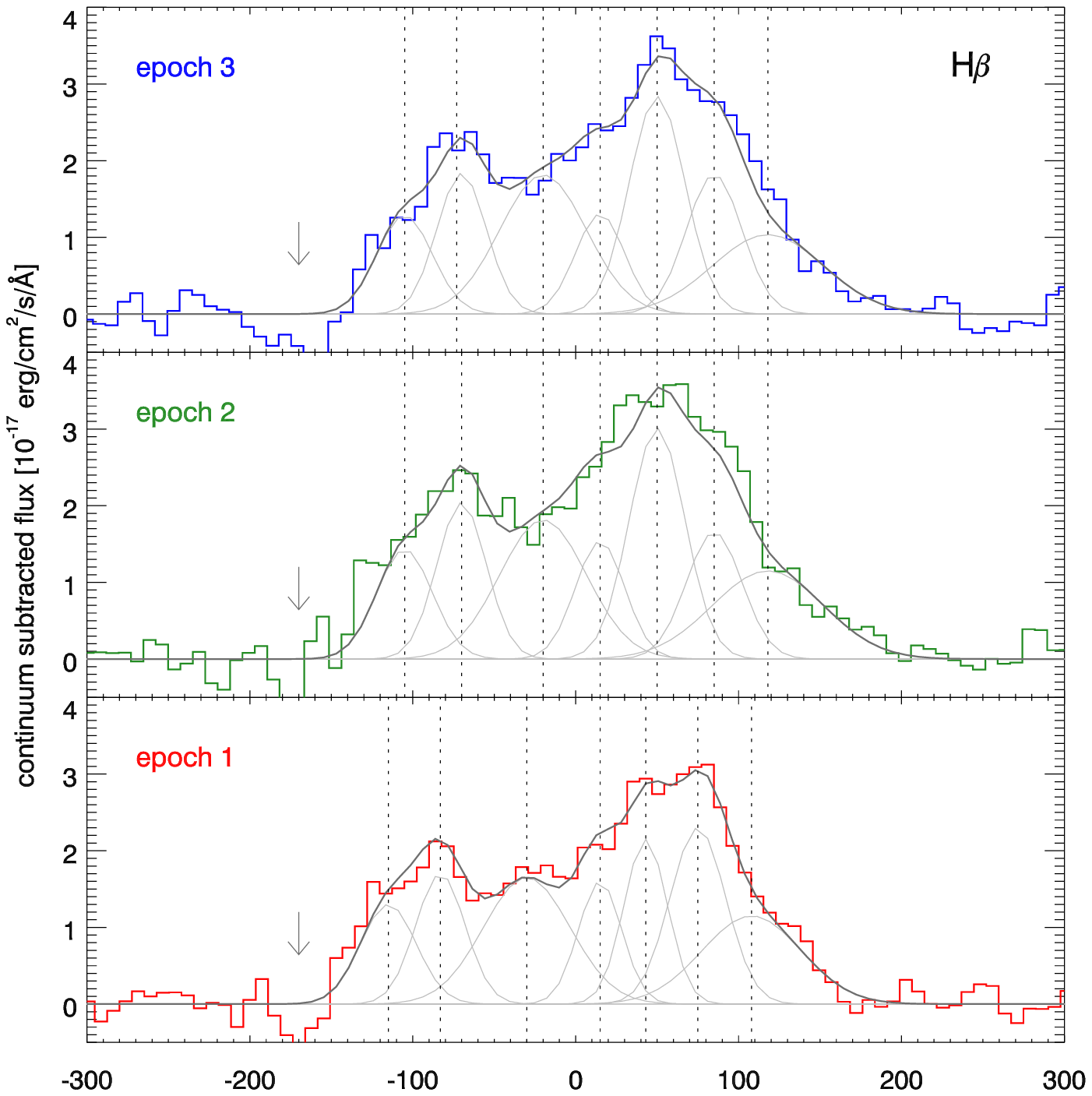}
    \includegraphics[width=6.5cm]{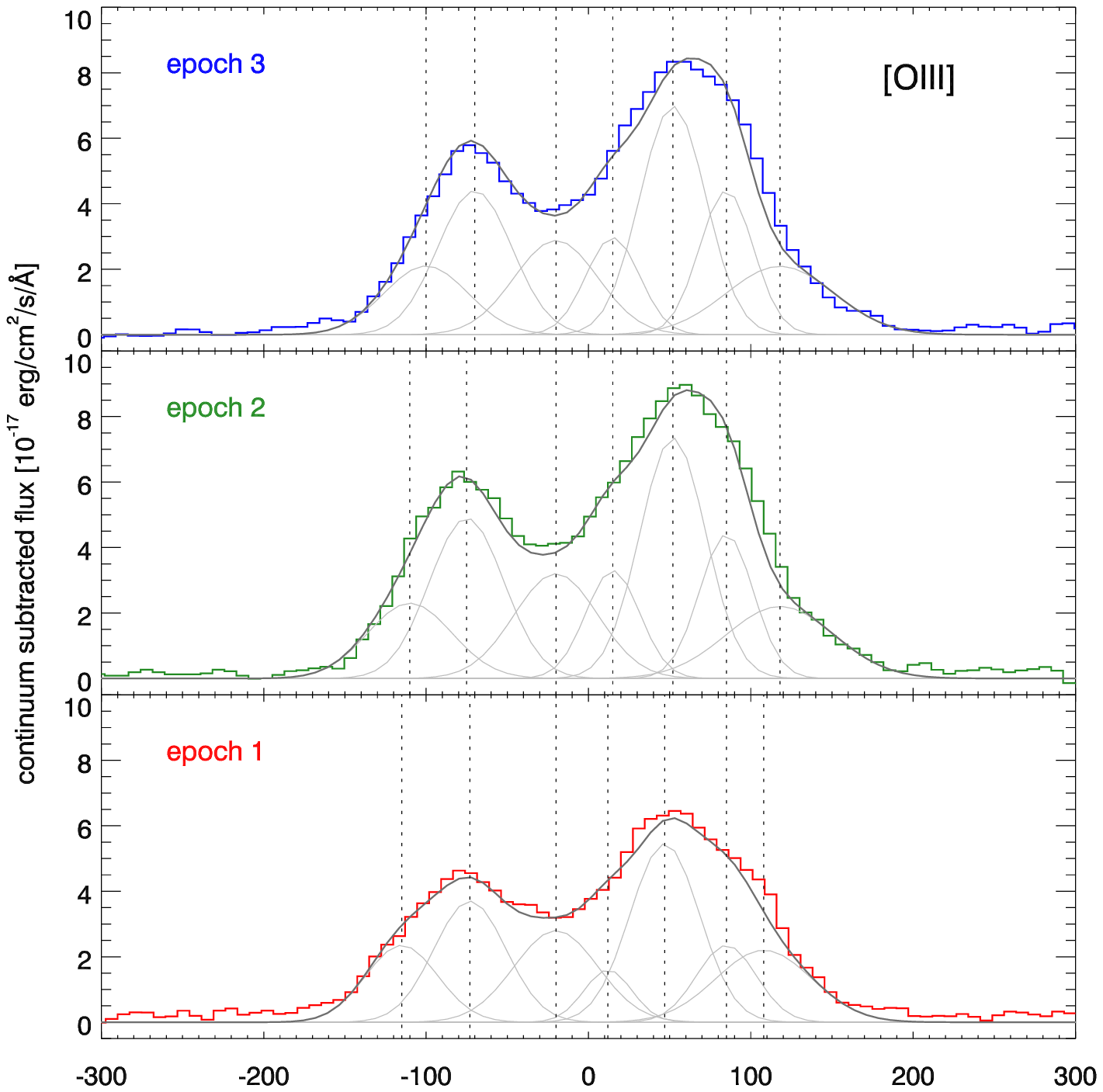} 
    \includegraphics[width=6.5cm]{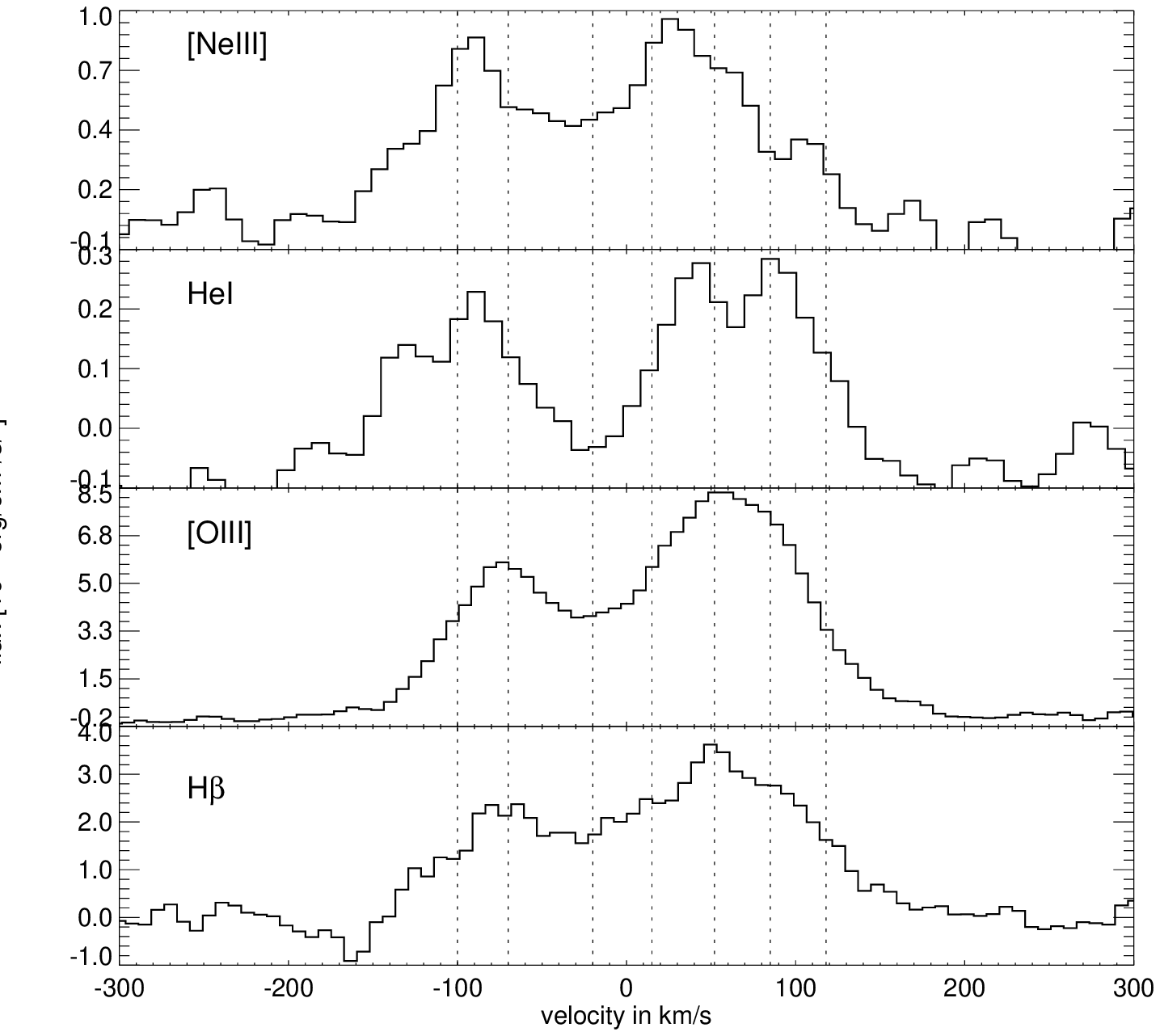}
      \caption{Gaussian fits to the H$\beta$ (top) and [OIII] (middle) emission lines at the three epochs of X-shooter spectra analysed here. The arrows in the plot of H$\beta$ indicate a possible stellar absorption component that we do not fit since the flux goes to negative values. The bottom panel shows [NeIII] and HeI emission lines from the third epoch compared to H$\beta$/[OIII] and the main (narrow) components fitted in H$\beta$ indicated for comparison.}
         \label{fig:HbO3fit}
   \end{figure}

The resolution of X-shooter and the simultaneous detection of emission and absorption lines in a single spectrum allows us to have a closer look at the galaxy kinematics. The structure in the emission lines with two main knots is visible in all three epochs but with slightly different relative strength and spatial extent/orientation due to the different orientation of the slit (see also Fig.\ref{Fig:Em}). In imaging we do not see any hint of a double component, only a small elongation in N-S direction, hence the two regions are aligned close to our line-of-sight. As expected, the components in epoch 2 \& 3 are very similar to each other since their slit orientation is very similar and they are also the two slits most aligned with the orientation of the galaxy. Epoch 1 differs somewhat due to an orientation almost perpendicular to the orientation of the galaxy. As we concluded above, the very similar abundances in both knots implies that the two main components are likely just two large SF regions in the host and not a pair of dwarf galaxies in close interaction. 

We fit several Gaussian components to the strongest emission lines of H$\beta$ and \ion{[O}{III]} as proxies for neutral and ionised gas. H$\alpha$ is in the NIR arm of X-shooter and has a lower resolution and S/N, hence we use H$\beta$ instead. In the following we name the two main emission components "blue" and "red blob" according to their relative line-of-sight velocities. The red blob is likely the SF region hosting the actual GRB and has stronger emission lines. In Fig. \ref{fig:HbO3fit} we plot the fit to the H$\beta$ and [OIII] lines which we describe below as well as the two lines together with He I and [Ne III] emission to compare their relative kinematics.  We do not perform component fits for the latter two due to their much lower S/N.

The blue blob is largely dominated by one single emission component, while the red blob can be divided in three narrow components with FWHM between 30 and 40 km\,s$^{-1}$. In the red wing of the red blob, an additional broad component with a $\sigma$ of 80 km\,s$^{-1}$ is needed while the situation is less clear in the blue wing, at least for H$\beta$ which could be affected by stellar absorption as can be seen in the spectra, however, the S/N is too low and the spectra go below zero, making an analysis of this possible absorption complicated. [OIII], however, does show an equivalent broad component in the blue wing of the blue blob and possibly even some emission beyond that. Broad components, interpreted as indications for outflows, have been found for the highly SF ''Green Pea'' galaxies \citep{amo12}, albeit with higher FWHM. We also need an additional component in the ''bridge'' between the two blobs, which we fit with a single broader component, probably a blend of several narrow components, but our resolution does not allow us to disentangle those. The components of epoch 2+3 in both H$\beta$ and [OIII] are slightly shifted to the red, especially in the blue blob and components change relative strength in the red blob. Epoch 1 is hence probably probing some blue shifted gas in the blue blob which is outside the main N-S orientation of the host. 

Another surprising observation can be done comparing [NeIII] and He I with H$\beta$ and [OIII], shown in Fig. \ref{fig:HbO3fit}. For both [NeIII] and HeI, the blue blob is clearly blue shifted compared to the blue blob in [OIII] and H$\beta$. For HeI, the components of the red blob match more or less to the ones of [OIII] and H$\beta$, however, the relative strength is different, with the main emission component of [OIII] and H$\beta$ being weaker in He I. For [NeIII], the main component in the red blob is clearly blueshifted compared to H$\beta$. As mentioned in Sect. \ref{sec:emlines}, [NeIII] and [OIII] are isoelectric equivalents, have similar ionisation energies, and are both produced in the $\alpha$ process in massive stars, hence they are usually expected and have been observed to be co-spatial and their ratio is constant with metallicity \citep{izo11}. One possibility to change their observed strength is extinction, which affects OIII more than NeIII, however, H$\alpha$ shows the same shape as H$\beta$ in our spectra, ruling out this possibility. Another way is an enhanced production of the Ne$^{22}$ isotope in e.g. WR stars, which has been observed in z$\sim$2 galaxies \citep{zei15}. However, the flux ratio of [NeIII]/[OIII] is consistent with the normal ratio observed in SDSS galaxies, only the position of the line is different, rather pointing to a difference in e.g. ionisation processes or differential outflows than a large difference in the element production. He I is only found in very young populations but dependent on temperature and density, hence the different relative strength compared to H$\beta$ might simply be a reflection of different conditions throughout the host.  

For the host of GRB 100418A we can furthermore compare absorption components from cold gas in the GRB sightline with the emitting gas from the galaxy. As mentioned in Sect. \ref{Sect:abs}, the absorbing gas probably has a different composition in the different components. Also, absorption and emission components mostly do not trace each other: only the two components with the highest redshift are consistent with the component of the ''bridge'' and one of the components in the red blob. The other absorption components span a much larger velocity range to the blue of the host complex with a total span of $>$500 km\,s$^{-1}$. The largest velocities are reached in \ion{Fe}{II} and \ion{Mg}{I}\&\ion{Mg}{II} while the equally strong \ion{Ca}{II} only shows absorption within the velocity range of the host emission (see Fig. \ref{Fig:Em}). \ion{Mg}{II} is often detected in extended halos of galaxies at large impact factors \citep[see e.g.][]{chu05,wer12} hence the absorption could correspond to cold gas around the galaxy complex or be indicative of outflows from the SF region, in addition to possibly observed outflows in the emitting gas (see above). A similar difference in velocity span between absorption and emission had been detected for the host of GRB 030329 \citep{tho07} and this was interpreted as a starburst wind driving metal rich gas out of the very low mass host.


\section{Conclusions}
\label{con}

In this paper we present a complete data set on GRB\,100418A and its host galaxy, at a redshift of $z=0.6239$. Observations include data ranging from radio to X-rays, with particularly high quality VLT/X-shooter spectroscopy, obtained at three epochs of the afterglow evolution. These spectra show a large number of both absorption and emission features that allow us to perform an extensive analysis of the host galaxy of this event.

The afterglow is characterised by a very rapid rebrightening in the optical, approximately 2.4 hr after the burst onset, with a brightness increase of $3.4\pm0.5$ mag. Similar very steep rebrightenings have been previously observed in other GRB afterglows during the first day after the burst. While the overall broadband evolution of the afterglow can be satisfactorily explained within the standard fireball model, the spectral slope of $\alpha_{rise}=5.4\pm1.1$ during the rebrightening rise cannot, and requires another interpretation. Although a complete, satisfactory explanation for these steep light curve rebrightenings is still lacking our observations favour an intense and delayed central engine activity.

Late analysis of the light curve indicates the presence of a faint supernova with an absolute magnitude of $M_V=-18.49\pm0.13$ mag at the time when one would expect the supernova to reach its maximum. The bright host galaxy, which is over 2.5 magnitudes brighter than the supernova at its peak time complicates this analysis. The SN associated with GRB\,100418A is among the faintest GRB-supernova detected to date. We caution that the limited available data at the time of the supernova does not allow us to make a full analysis of its evolution and characteristics. Such a faint supernova is indicative of the varied characteristics of events of this type.

The host of GRB 100418A is a prototypical GRB host at $z\sim0.6$, but with higher SFR and SSFR and a metallicity (12+log(O/H)$=$8.55) consistent with the local mass-metallicity relation.  While it appears very compact in imaging, the spectral resolution of X-shooter and the detection of a large number of bright emission lines allow us to determine both its emission- and absorption-line kinematics. The compact galaxy resolves into two distinct knots with similar properties suggesting that they are two large SF regions of a dwarf galaxy and not a system in a late stage of merging. The absorption lines span a velocity range of $>$500\,km\,s$^{-1}$, while the hot emitting gas only reaches a total velocity of $\sim$300\,km\,s$^{-1}$. Both broad wings in the emission lines and the high velocities of the absorption lines point to outflows of metal-enriched gas from the intense star formation in the host, similar to what has been observed for green pea galaxies and some GRB hosts.

\begin{acknowledgements}
AdUP, CCT, KB, DAK, LI, and ZC acknowledge support from the Spanish Ministry of Economy and Competitivity under grant number AYA 2014-58381-P; in addition, AdUP and CCT from Ram\'on y Cajal fellowships (RyC-2012-09975 and RyC-2012-09984). DAK also acknowledges financial support from Juan de la Cierva Incorporaci\'on fellowship IJCI-2015-26153. RF acknowledges support from European Regional Development Fund-Project "Engineering applications of microworld physics" (No. CZ.02.1.01/0.0/0.0/16\_019/0000766). IB, IKh, RB, and SM acknowledge TUBITAK, IKI, KFU, and AST for partial support in using RTT150. This work was partially funded by the subsidy 3.6714.2017/8.9 allocated to Kazan Federal University for the state assignment in the sphere of scientific activities.
TK acknowledges support from the DFG cluster of excellence `Origin and Structure of the Universe'. PS  acknowledges support through the Sofja Kovalevskaja Award from the Alexander von Humboldt Foundation of Germany. RSR acknwoledges support from ASI (Italian Space Agency) through Contract n. 2015-046-R.0 and from European Union Horizon 2020 Programme under the AHEAD project (grant agreement n. 654215). The Cosmic Dawn Center is funded by the DNRF.

Based on observations made with the Gran Telescopio Canarias (GTC), installed in the Spanish Observatorio del Roque de los Muchachos of the Instituto de Astrof\' isica de Canarias, in the island of La Palma. 
The Submillimeter Array is a joint project between the Smithsonian Astrophysical Observatory and the Academia Sinica Institute of Astronomy and Astrophysics and is funded by the Smithsonian Institution and the Academia Sinica. This work is based on observations carried out under project number U051 with the 
IRAM Plateau de Bure Interferometer. IRAM is supported by INSU/CNRS (France), MPG (Germany), and IGN (Spain).
Some of the data presented herein were obtained at the W. M. Keck
Observatory, which is operated as a scientific partnership among the
California Institute of Technology, the University of California, and the
National Aeronautics and Space Administration. The Observatory was made
possible by the generous financial support of the W. M. Keck Foundation.
This work is based on observations made with the Spitzer Space
Telescope, which is operated by the Jet Propulsion Laboratory,
California Institute of Technology under a contract with NASA. Support
for this work was provided by NASA through an award issued by JPL/Caltech.
The UKIRT data were pipeline processed at the 
Cambridge Astronomical Survey Unit, and
are archived at the Wide Field Astronomy
Unit at the Royal Observatory Edinburgh
This work made use of data supplied by the UK Swift Science Data Centre at the University of Leicester.

\end{acknowledgements}

\bibliographystyle{aa}
\bibliography{100418A}


\newpage\clearpage
\onecolumn
\begin{appendix}

\section{}




\begin{longtable}{c c c c c}   
\caption{Optical/NIR observations. Magnitudes are in the AB system, without correction for galactic extinction, whereas the flux densities have been corrected for it.
\label{table:vis}}\\
\hline\hline                 
T-T$_0$ 			& Telescope 	& Band	    & Magnitude         & Flux density          \\    
(days)     			&               &     	    &			        & ($\mu$Jy) 	        \\
\hline                        
\endfirsthead
\caption{continued.}\\
\hline\hline 
T-T$_0$ 			& Observatory   & Band	    & Magnitude         & Flux                  \\     
(days)     			&               &           &			        & ($\mu$Jy)             \\
\hline                        	
\endhead
\hline
\endfoot
        0.07464 & {\it Swift}/UVOT & {\it uvw2}   & 22.19$\pm$0.25    &    7.92$\pm$2.06    \\
        6.24391 & {\it Swift}/UVOT & {\it uvw2}   & 24.03$\pm$0.30    &    1.45$\pm$0.47    \\
        8.33782 & {\it Swift}/UVOT & {\it uvw2}   & 23.54$\pm$0.22    &    2.28$\pm$0.52    \\
\hline 
        0.01117 & {\it Swift}/UVOT & {\it uvm2}   & 21.56$\pm$0.34    &   15.47$\pm$5.71    \\
        0.06275 & {\it Swift}/UVOT & {\it uvm2}   & 21.78$\pm$0.22    &   12.65$\pm$2.89    \\
        0.07939 & {\it Swift}/UVOT & {\it uvm2}   & 21.55$\pm$0.20    &   15.65$\pm$3.20    \\
        5.25524 & {\it Swift}/UVOT & {\it uvm2}   & 23.07$\pm$0.19    &    3.86$\pm$0.73    \\
        7.89360 & {\it Swift}/UVOT & {\it uvm2}   & 23.59$\pm$0.23    &    2.39$\pm$0.56    \\
\hline 
        0.01145 & {\it Swift}/UVOT & {\it uvw1}   & 21.42$\pm$0.32    &   14.90$\pm$5.19    \\
        0.06513 & {\it Swift}/UVOT & {\it uvw1}   & 22.01$\pm$0.29    &    8.61$\pm$2.64    \\
        0.08176 & {\it Swift}/UVOT & {\it uvw1}   & 21.77$\pm$0.29    &   10.72$\pm$3.27    \\
        2.59007 & {\it Swift}/UVOT & {\it uvw1}   & 22.25$\pm$0.11    &    6.89$\pm$0.70    \\
        6.43383 & {\it Swift}/UVOT & {\it uvw1}   & 23.52$\pm$0.31    &    2.14$\pm$0.72    \\
        8.35208 & {\it Swift}/UVOT & {\it uvw1}   & 23.68$\pm$0.30    &    1.85$\pm$0.58    \\
\hline 
        0.00187 & {\it Swift}/UVOT & {\it white}& 21.46$\pm$0.15    &   12.57$\pm$1.91    \\
        0.01035 & {\it Swift}/UVOT & {\it white}& 21.67$\pm$0.17    &   10.43$\pm$1.73    \\
        0.07226 & {\it Swift}/UVOT & {\it white}& 21.14$\pm$0.11    &   16.98$\pm$1.73    \\
        0.60178 & {\it Swift}/UVOT & {\it white}& 20.18$\pm$0.02    &   40.85$\pm$0.65    \\
        1.01772 & {\it Swift}/UVOT & {\it white}& 20.67$\pm$0.02    &   26.18$\pm$0.58    \\
        1.76396 & {\it Swift}/UVOT & {\it white}& 21.30$\pm$0.03    &   14.60$\pm$0.39    \\
        2.79714 & {\it Swift}/UVOT & {\it white}& 21.83$\pm$0.04    &    8.95$\pm$0.36    \\
        3.54998 & {\it Swift}/UVOT & {\it white}& 22.20$\pm$0.04    &    6.35$\pm$0.22    \\
        4.43120 & {\it Swift}/UVOT & {\it white}& 22.26$\pm$0.05    &    6.03$\pm$0.30    \\
        5.49919 & {\it Swift}/UVOT & {\it white}& 22.41$\pm$0.07    &    5.24$\pm$0.33    \\
        7.27573 & {\it Swift}/UVOT & {\it white}& 22.40$\pm$0.10    &    5.32$\pm$0.50    \\
        9.32466 & {\it Swift}/UVOT & {\it white}& 23.01$\pm$0.09    &    3.01$\pm$0.26    \\
       10.77969 & {\it Swift}/UVOT & {\it white}& 23.15$\pm$0.06    &    2.65$\pm$0.15    \\
       13.03649 & {\it Swift}/UVOT & {\it white}& 23.11$\pm$0.05    &    2.77$\pm$0.14    \\
\hline 
        2.3256  & 10.4mGTC/OSIRIS  & {\it u$^\prime$}     & 21.55$\pm$0.05   &   11.58$\pm$0.53    \\ 
        4.2138  & 10.4mGTC/OSIRIS  & {\it u$^\prime$}     & 21.68$\pm$0.06   &   10.27$\pm$0.57                   \\ 
        90.1587 & 10.4mGTC/OSIRIS  & {\it u$^\prime$}     & 23.17$\pm$0.06   &    2.60$\pm$0.14               \\ 
\hline
        0.07524 & {\it Swift}/UVOT & {\it u}     & 20.98$\pm$0.19 &   19.86$\pm$3.74  \\
        5.76062 & {\it Swift}/UVOT & {\it u}     & 22.32$\pm$0.36 &    5.78$\pm$2.28  \\
        7.94797 & {\it Swift}/UVOT & {\it u}     & 23.03$\pm$0.23 &    2.99$\pm$0.72  \\
\hline 
        0.06988 & {\it Swift}/UVOT & {\it b}     & 20.27$\pm$0.20 &   35.81$\pm$7.36  \\
        6.90466 & {\it Swift}/UVOT & {\it b}     & 22.27$\pm$0.34 &    5.72$\pm$2.14  \\
        8.55971 & {\it Swift}/UVOT & {\it b}     & 22.24$\pm$0.36 &    5.83$\pm$2.26  \\
        1067.3543 & 10.0mKeck/LRIS & {\it B}     & 23.23$\pm$0.10 &    2.63$\pm$0.22               \\
\hline 
	1.50975  & 8.2mVLT/X-shooter & {\it g$^\prime$}    & 20.45$\pm$0.09  &   29.95$\pm$2.49   \\ 
        2.3355  &10.4mGTC/OSIRIS& {\it g$^\prime$}     & 21.05$\pm$0.02 &   17.23$\pm$0.32               \\
        4.2347  &10.4mGTC/OSIRIS& {\it g$^\prime$}     & 21.56$\pm$0.06 &   10.77$\pm$0.60                \\
        17.7392 & 10.0mKeck/LRIS& {\it g$^\prime$}     & 22.71$\pm$0.06 &    3.74$\pm$0.21               \\
       28.3172  &10.4mGTC/OSIRIS& {\it g$^\prime$}     & 22.94$\pm$0.04 &    3.02$\pm$0.11               \\
       90.1744  &10.4mGTC/OSIRIS& {\it g$^\prime$}     & 23.01$\pm$0.04 &    2.83$\pm$0.10               \\
       2647.101 &10.4mGTC/OSIRIS& {\it g$^\prime$}     & 22.99$\pm$0.09 &    2.87$\pm$0.24               \\
\hline           
        6.5145  & 8.2mVLT/FORS2 & {\it V}     & 21.86$\pm$0.04 &    7.87$\pm$  0.29        \\
\hline           
	0.34872  & 8.2mVLT/X-shooter & {\it r$^\prime$}    & 18.79$\pm$0.04  &   128.94$\pm$4.75  \\ 
	1.44281  & 8.2mVLT/X-shooter & {\it r$^\prime$}    & 19.85$\pm$0.07  &   48.57$\pm$3.13   \\ 
	1.50698  & 8.2mVLT/X-shooter & {\it r$^\prime$}    & 19.98$\pm$0.08  &   43.09$\pm$3.18   \\ 
        2.3394  &10.4mGTC/OSIRIS& {\it r$^\prime$}     & 20.59$\pm$0.04 &   24.57$\pm$0.91               \\  
	2.46029  & 8.2mVLT/X-shooter & {\it r$^\prime$}    & 20.66$\pm$0.09  &   23.04$\pm$1.91   \\ 
        4.2391  &10.4mGTC/OSIRIS& {\it r$^\prime$}     & 21.09$\pm$0.04 &   15.50$\pm$0.57               \\ 
       28.3049  &10.4mGTC/OSIRIS& {\it r$^\prime$}     & 22.23$\pm$0.04 &    5.42$\pm$0.20              \\ 
        90.1826 &10.4mGTC/OSIRIS& {\it r$^\prime$}     & 22.34$\pm$0.04 &    4.90$\pm$0.18              \\ 
       2646.1457&10.4mGTC/OSIRIS& {\it r$^\prime$}     & 22.43$\pm$0.05 &    4.51$\pm$0.21              \\ 
\hline           
        0.02276 & RTT150 & {\it R$_C$}            & 20.55$\pm$0.05 &   25.33$\pm$1.17  \\
        0.06085 & RTT150 & {\it R$_C$}            & 21.44$\pm$0.26 &   11.16$\pm$3.08  \\
        0.09378 & RTT150 & {\it R$_C$}            & 19.07$\pm$0.04 &   99.01$\pm$3.65  \\
        0.09825 & RTT150 & {\it R$_C$}            & 18.97$\pm$0.09 &  108.57$\pm$9.00  \\
        0.10723 & RTT150 & {\it R$_C$}            & 18.87$\pm$0.18 &  119.04$\pm$20.83  \\
        0.11173 & RTT150 & {\it R$_C$}            & 18.71$\pm$0.03 &  137.94$\pm$3.81  \\
        0.11637 & RTT150 & {\it R$_C$}            & 18.73$\pm$0.03 &  135.42$\pm$3.74  \\
        1.14583 & RTT150 & {\it R$_C$}            & 19.33$\pm$0.09 &   77.93$\pm$6.46  \\
        2.14583 & RTT150 & {\it R$_C$}            & 20.46$\pm$0.07 &   27.52$\pm$1.77  \\
        2.15417 & RTT150 & {\it R$_C$}            & 20.44$\pm$0.09 &   28.03$\pm$2.32  \\
        2.16250 & RTT150 & {\it R$_C$}            & 20.49$\pm$0.05 &   26.77$\pm$1.23  \\
        2.20833 & RTT150 & {\it R$_C$}            & 20.50$\pm$0.03 &   26.53$\pm$0.73  \\
        5.12917 & RTT150 & {\it R$_C$}            & 20.95$\pm$0.20 &   17.53$\pm$3.55  \\
        6.5172  & 8.2mVLT/FORS & {\it R$_C$}      & 21.52$\pm$0.06 &   10.36$\pm$0.57    \\
        9.16667 & RTT150 & {\it R$_C$}            & 21.66$\pm$0.09 &    9.11$\pm$0.76  \\
       13.06250 & RTT150 & {\it R$_C$}            & 21.98$\pm$0.08 &    6.79$\pm$0.50  \\
       14.10417 & RTT150 & {\it R$_C$}            & 21.98$\pm$0.05 &    6.79$\pm$0.31  \\
       15.13750 & RTT150 & {\it R$_C$}            & 21.96$\pm$0.03 &    6.91$\pm$0.19  \\
       16.10417 & RTT150 & {\it R$_C$}            & 21.95$\pm$0.03 &    6.98$\pm$0.19  \\
       17.7395  & 10.0mKeck/LRIS & {\it R$_C$}    & 22.02$\pm$0.06 &    6.53$\pm$0.36     \\
       18.14583 & RTT150 & {\it R$_C$}            & 22.11$\pm$0.06 &    6.02$\pm$0.33  \\
       21.12500 & RTT150 & {\it R$_C$}            & 22.11$\pm$0.05 &    6.02$\pm$0.28  \\
       22.10417 & RTT150 & {\it R$_C$}            & 22.12$\pm$0.05 &    5.97$\pm$0.27  \\
       25.6361  & 8.2mSubaru/FOCAS & {\it R$_C$}  & 22.31$\pm$0.05 &    5.00$\pm$0.23  \\
       57.07083 & RTT150 & {\it R$_C$}            & 22.24$\pm$0.06 &    5.34$\pm$0.30  \\
      106.96389 & RTT150 & {\it R$_C$}            & 22.31$\pm$0.05 &    5.01$\pm$0.23  \\
\hline            
	 1.51242  & 8.2mVLT/X-shooter & {\it i$^\prime$}    & 19.63$\pm$0.07  &   53.65$\pm$3.69   \\ 
        2.3419  &10.4mGTC/OSIRIS& {\it i$^\prime$}     & 20.22$\pm$0.03 &   33.24$\pm$0.92               \\ 
        4.2435  &10.4mGTC/OSIRIS& {\it i$^\prime$}     & 20.66$\pm$0.03 &   22.16$\pm$0.61               \\ 
       28.3131  &10.4mGTC/OSIRIS& {\it i$^\prime$}     & 21.86$\pm$0.04 &    7.34$\pm$0.27              \\ 
        90.1961 &10.4mGTC/OSIRIS& {\it i$^\prime$}     & 22.07$\pm$0.04 &    6.04$\pm$0.22              \\ 
       2646.1187&10.4mGTC/OSIRIS& {\it i$^\prime$}     & 21.90$\pm$0.06 &    7.07$\pm$0.39              \\ 
\hline
        6.5197  & 8.2mVLT/FORS  & {\it I$_C$}     & 21.06$\pm$0.03 &    15.14$\pm$0.42 \\
        15.3697 & 8.2mVLT/FORS  & {\it I$_C$}     & 21.40$\pm$0.03 &    11.07$\pm$0.31 \\
       1067.7231&10.0mKeck/LRIS & {\it I$_C$}     & 21.98$\pm$0.06 &    6.49$\pm$0.36 \\
\hline
        2.3460  &10.4mGTC/OSIRIS& {\it z$^\prime$}     & 20.07$\pm$0.05 &   37.05$\pm$1.71               \\ 
        4.2504  &10.4mGTC/OSIRIS& {\it z$^\prime$}     & 20.57$\pm$0.05 &   23.38$\pm$1.08               \\ 
       90.2058  &10.4mGTC/OSIRIS& {\it z$^\prime$}     & 22.02$\pm$0.07 &    6.15$\pm$0.40              \\ 
      2647.1253 &10.4mGTC/OSIRIS& {\it z$^\prime$}     & 22.03$\pm$0.07 &    6.09$\pm$0.39              \\ 
\hline
        1.56641 &3.8mUKIRT/WFCam& {\it J}     & 19.07$\pm$0.04 &   89.70$\pm$3.30  \\
        1.57644 &3.8mUKIRT/WFCam& {\it J}     & 19.03$\pm$0.04 &   93.07$\pm$3.43  \\
        5.24920 & 3.5mCAHA/O2000& {\it J}     & 20.57$\pm$0.08 &   22.51$\pm$1.66  \\
       13.15972 & 3.5mCAHA/O2000& {\it J}     & 21.81$\pm$0.25 &    7.18$\pm$1.85  \\
       41.13148 & 3.5mCAHA/O2000& {\it J}     & 21.94$\pm$0.13 &    6.37$\pm$0.82  \\
\hline
        1.56928 &3.8mUKIRT/WFCam& {\it H}     & 18.68$\pm$0.03 &  126.24$\pm$3.49  \\
        1.57978 &3.8mUKIRT/WFCam& {\it H}     & 18.68$\pm$0.03 &  126.24$\pm$3.49  \\
        8.11896 & 3.5mCAHA/O2000& {\it H}     & 21.01$\pm$0.16 &   14.82$\pm$2.32  \\
\hline
        1.57269 &3.8mUKIRT/WFCam& {\it K$_S$} & 18.32$\pm$0.02 &  174.10$\pm$3.21  \\
        1.58329 &3.8mUKIRT/WFCam& {\it K$_S$} & 18.39$\pm$0.03 &  163.23$\pm$4.51   \\
        7.13634 & 3.5mCAHA/O2000 & {\it K$_S$} & 20.78$\pm$0.20 &   17.98$\pm$3.64  \\
        8.15765 & 3.5mCAHA/O2000 & {\it K$_S$} & 20.71$\pm$0.19 &   19.18$\pm$3.71  \\
        1007.3888 & 4.2mWHT/LIRIS & {\it K$_S$}& 21.57$\pm$0.21 &    8.71$\pm$1.69  \\
\hline
        408.5937  & {\it Spitzer}& {\it 3.6 $\mu$m}& 21.98$\pm$0.10 &    5.86$\pm$0.57   \\
\hline
\end{longtable}
\twocolumn


\begin{figure*}
\centering
\includegraphics[width=14cm]{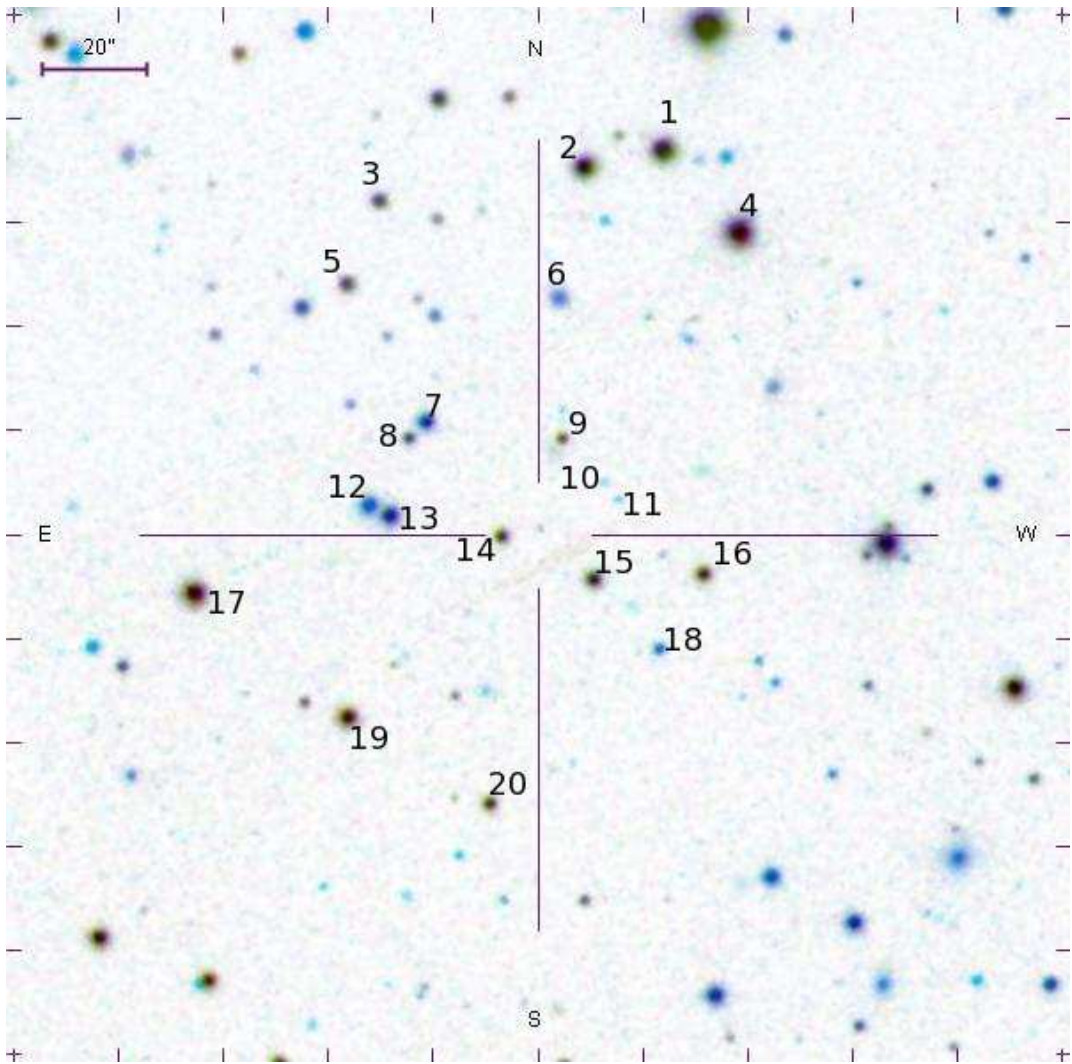}
\caption{Calibration stars used for the optical photometry.}
\label{fcfig}
\end{figure*}

\begin{table*}
\caption{Calibration stars used for photometry in Sloan bands (AB) and Johnson/Cousins (Vega), as indicated in Fig.~\ref{fcfig}. The transformation from Sloan to Johnson/Cousins was done using the \citet{jes05} transformations.}             
\label{table:fc}      
\centering          
\begin{tabular}{c c c c c c}     
\hline\hline       
Star & \textit{u} & \textit{g} &\textit{r} & \textit{i} & \textit{z} \\ 
\hline                    
1	& 18.40$\pm$0.02	& 17.07$\pm$0.01	& 16.53$\pm$0.01	& 16.33$\pm$0.01	& 16.24$\pm$0.01	\\
2	& 18.94$\pm$0.02	& 17.47$\pm$0.01	& 16.93$\pm$0.01	& 16.73$\pm$0.01	& 16.62$\pm$0.01	\\
3	& 21.20$\pm$0.11	& 19.48$\pm$0.01	& 18.82$\pm$0.01	& 18.56$\pm$0.01	& 18.42$\pm$0.03	\\
4	& 17.35$\pm$0.01	& 15.82$\pm$0.01	& 15.27$\pm$0.01	& 15.09$\pm$0.01	& 15.00$\pm$0.01	\\
5	& 20.99$\pm$0.09	& 19.36$\pm$0.01	& 18.74$\pm$0.01	& 18.49$\pm$0.01	& 18.31$\pm$0.03	\\
6	& ----				& 20.80$\pm$0.04	& 19.21$\pm$0.02	& 18.65$\pm$0.01	& 18.29$\pm$0.05	\\
7	& 21.63$\pm$0.15	& 19.15$\pm$0.01	& 17.82$\pm$0.01	& 17.27$\pm$0.01	& 16.97$\pm$0.01	\\
8	& 22.06$\pm$0.21	& 20.16$\pm$0.02	& 19.35$\pm$0.01	& 18.98$\pm$0.01	& 18.80$\pm$0.04	\\
9	& 21.13$\pm$0.10	& 19.87$\pm$0.02	& 19.54$\pm$0.02	& 19.33$\pm$0.02	& 19.34$\pm$0.07	\\
10	& ---				& ---				& 22.69$\pm$0.20	& 21.60$\pm$0.11	& ---				\\
11	& ---				& 23.79$\pm$0.33	& 22.27$\pm$0.13	& 21.09$\pm$0.07	& 20.95$\pm$0.28	\\
12	& 21.78$\pm$0.17	& 19.01$\pm$0.01	& 17.62$\pm$0.01	& 16.91$\pm$0.01	& 16.67$\pm$0.01	\\
13	& 20.82$\pm$0.08	& 18.31$\pm$0.01	& 17.07$\pm$0.01	& 16.59$\pm$0.01	& 16.31$\pm$0.01	\\
14	& 20.15$\pm$0.05	& 19.00$\pm$0.01	& 18.58$\pm$0.01	& 18.37$\pm$0.01	& 18.35$\pm$0.03	\\
15	& 19.74$\pm$0.04	& 18.34$\pm$0.01	& 17.77$\pm$0.01	& 17.54$\pm$0.01	& 17.38$\pm$0.02	\\
16	& 19.51$\pm$0.03	& 18.52$\pm$0.01	& 18.11$\pm$0.01	& 17.92$\pm$0.01	& 17.84$\pm$0.02	\\
17	& 17.88$\pm$0.01	& 16.54$\pm$0.01	& 16.06$\pm$0.01	& 15.76$\pm$0.01	& 15.79$\pm$0.01	\\
18	& ---				& 20.97$\pm$0.03	& 19.62$\pm$0.02	& 18.94$\pm$0.01	& 18.68$\pm$0.04	\\
19	& 18.65$\pm$0.02 	& 17.54$\pm$0.01	& 17.13$\pm$0.01	& 16.97$\pm$0.01	& 16.88$\pm$0.01	\\
20	& 20.21$\pm$0.05	& 18.95$\pm$0.01	& 18.54$\pm$0.01	& 18.37$\pm$0.01	& 18.33$\pm$0.03	\\
\hline   
Star & \textit{U} & \textit{B} &\textit{V} & \textit{R$_C$} & \textit{I$_C$} \\ 
\hline                    
1	& 17.65$\pm$0.04	& 17.49$\pm$0.03	& 16.74$\pm$0.02	& 16.30$\pm$0.03	& 15.89$\pm$0.02	\\
2	& 18.16$\pm$0.04	& 17.89$\pm$0.03	& 17.14$\pm$0.02	& 16.70$\pm$0.03	& 16.29$\pm$0.02	\\
3	& 21.41$\pm$0.12	& 19.95$\pm$0.03	& 19.08$\pm$0.02	& 18.58$\pm$0.03	& 18.11$\pm$0.03	\\
4	& 16.56$\pm$0.04	& 16.24$\pm$0.03	& 15.49$\pm$0.02	& 15.07$\pm$0.03	& 14.68$\pm$0.02	\\
5	& 20.30$\pm$0.10	& 19.81$\pm$0.03	& 18.98$\pm$0.02	& 18.49$\pm$0.03	& 18.03$\pm$0.03	\\
7	& 20.93$\pm$0.16	& 19.87$\pm$0.03	& 18.36$\pm$0.02	& 17.54$\pm$0.03	& 16.77$\pm$0.02	\\
8	& 21.29$\pm$0.21	& 20.69$\pm$0.04	& 19.67$\pm$0.02	& 19.05$\pm$0.03	& 18.47$\pm$0.04	\\
9	& 21.31$\pm$0.11	& 20.21$\pm$0.04	& 19.67$\pm$0.02	& 19.22$\pm$0.04	& 18.80$\pm$0.07	\\
12	& 21.04$\pm$0.17	& 19.76$\pm$0.03	& 18.18$\pm$0.02	& 17.19$\pm$0.03	& 16.27$\pm$0.02	\\
13	& 20.08$\pm$0.09	& 19.00$\pm$0.03	& 17.57$\pm$0.02	& 16.83$\pm$0.03	& 16.14$\pm$0.02	\\
14	& 19.39$\pm$0.06	& 19.37$\pm$0.03	& 18.74$\pm$0.02	& 18.29$\pm$0.03	& 17.87$\pm$0.03	\\
15	& 18.98$\pm$0.06	& 18.77$\pm$0.03	& 17.99$\pm$0.02	& 17.24$\pm$0.03	& 17.08$\pm$0.02	\\
16	& 18.74$\pm$0.05	& 18.86$\pm$0.02	& 18.27$\pm$0.02	& 17.88$\pm$0.02	& 17.48$\pm$0.02	\\
17	& 17.10$\pm$0.04	& 16.94$\pm$0.03	& 16.25$\pm$0.02	& 15.70$\pm$0.03	& 15.19$\pm$0.02	\\
19	& 17.85$\pm$0.04 	& 17.88$\pm$0.03	& 17.29$\pm$0.02	& 16.93$\pm$0.03	& 16.56$\pm$0.02	\\
20	& 19.42$\pm$0.06	& 19.32$\pm$0.02	& 18.70$\pm$0.02	& 18.29$\pm$0.02	& 17.91$\pm$0.03	\\
\hline  
\end{tabular}
\end{table*}

\begin{table*}[!htb]
\caption{Host galaxy metallicities (top part of the table) and abundances converted into a base metallicity according to \cite{kew02} (KD02).
 \label{tabz}}
\centering                           
\begin{scriptsize}
\begin{tabular}{c c c c c c c c c} 
\hline\hline                
\textbf{Calibrator}	& \textbf{Lines (Support lines)}	& \textbf{Z$_{Flux 1}$}  & \textbf{Z$_{Flux 2}$}   & \textbf{Z$_{F1 +F2}$} &	 \textbf{Z$_{F Comb}$}  		\\
				&						& {\scriptsize \textbf{(12  + log10(O/H))}} & {\scriptsize \textbf{(12  + log10(O/H))}} & {\scriptsize \textbf{(12  + log10(O/H))}}  & {\scriptsize \textbf{(12  + log10(O/H))}}   \\
\hline
\multicolumn{6}{c}{EPOCH 1} \\  
\hline
D02$^{1}$              &     N2                                                     &    8.401     -0.277     +0.162        &  8.409     -0.276     +0.174     &  8.370     -0.266     +0.168   &   8.264     -0.243     +0.162    \\
Z94$^{2}$              &     R$_{23}$                                               &    8.497     -0.033     +0.076        &  8.688     -0.015     +0.028     &  8.621     -0.016     +0.025   &   8.604     -0.006     +0.008    \\
M91$^{3}$              &     R$_{23}$, O3O2 (N2, N2O2)                              &    8.266     -0.046     +0.256        &  8.614     -0.633     +0.035     &  8.553     -0.491     +0.015   &   8.105     -0.013     +0.010    \\
PP04 N2Ha$^{4}$        &     N2                                                     &    8.296     -0.140     +0.098        &  8.300     -0.138     +0.100     &  8.280     -0.130     +0.090   &   8.209     -0.102     +0.061    \\
PP04 O3N2$^{4}$        &     N2, O3/H$\beta$                                        &    8.259     -0.115     +0.058        &  8.263     -0.108     +0.058     &  8.246     -0.108     +0.056   &   8.193     -0.090     +0.047    \\
P10 ONS$^{5}$                &     [N II]$\lambda$6584/H$\beta$, O3/H$\beta$        &    8.452     -0.154     +0.085        &  8.540     -0.159     +0.094     &  8.514     -0.159     +0.090   &   8.408     -0.113     +0.072    \\
                       &     [O II]$\lambda$3727/H$\beta$, [S II]/H$\beta$,&&&&   \\                       
P10 ON$^{5}$                 &     N2, O3/H$\beta$, [O II]$\lambda$3727/H$\beta$    &    7.977     -0.374     +0.210        &  8.071     -0.387     +0.230     &  8.004     -0.386     +0.219   &   7.707     -0.287     +0.186    \\
M08 N2Ha$^{6}$         &     N2                                                     &    8.597     -0.273     +0.093        &  8.591     -0.259     +0.099     &  8.539     -0.266     +0.151   &   8.386     -0.226     +0.117    \\
M08 O3O2$^{6}$         &     O3O2                                                   &    8.709     -0.059     +0.027        &  8.575     -0.042     +0.021     &  8.615     -0.031     +0.019   &   8.626     -0.014     +0.010    \\
M13 O3N2$^{7}$         &     [N II]$\lambda$6584/H$\beta$, O3/H$\beta$              &    8.219     -0.077     +0.039        &  8.220     -0.074     +0.039     &  8.208     -0.073     +0.039   &   8.175     -0.060     +0.032    \\
M13 N2$^{7}$           &     [N II]$\lambda$6584/H$\beta$                           &    8.292     -0.160     +0.089        &  8.298     -0.165     +0.089     &  8.276     -0.161     +0.087   &   8.201     -0.131     +0.077    \\
KD02 N2O2$^{8}$        &     N2O2                                                   &    8.389     -0.747     +0.095        &  8.579     -0.194     +0.090     &  8.487     -0.200     +0.096   &   8.293     -0.733     +0.109    \\
KK04 N2Ha$^{9}$        &     N2, q, (N2O2)                                          &    8.616     -0.182     +0.099        &  8.687     -0.224     +0.126     &  8.625     -0.194     +0.118   &   8.478     -0.140     +0.097    \\
KK04 R23$^{9}$         &     R$_{23}$, q, (N2, N2O2)                                &    8.438     -0.040     +0.217        &  8.759     -0.573     +0.035     &  8.691     -0.433     +0.016   &   8.288     -0.010     +0.008    \\
KD02comb$^{9}$         &     COMBINED$^*$                                           &    8.342     -0.039     +0.130        &  8.554     -0.474     +0.111     &  8.450     -0.292     +0.126   &   8.196     -0.011     +0.009    \\
\hline
KK04         &     R$_{23}$, q, (N2, N2O2)    & 8.326 -0.038 +0.208  & 8.657 -0.628 +0.038  & 8.583 -0.464 +0.017  &   8.188 -0.009 +0.007    \\
Z94          &     R$_{23}$                   & 8.444 -0.043 +0.098  & 8.659 -0.015 +0.028  & 8.590 -0.017 +0.027  &   8.572 -0.007 +0.009    \\
M91          &     R$_{23}$, O3O2 (N2, N2O2)  & 7.737 -0.197 +1.098  & 8.682 -0.957 +0.053  & 8.580 -0.903 +0.028  &   6.889 -0.082 +0.063    \\
D02          &     N2                         & 8.551 -0.234 +0.137  & 8.558 -0.236 +0.149  & 8.526 -0.214 +0.135  &   8.448 -0.162 +0.108    \\
PP04O3N2     &     N2, O3/H$\beta$            & 8.532 -0.101 +0.051  & 8.536 -0.095 +0.051  & 8.521 -0.094 +0.049  &   8.476 -0.075 +0.039    \\
PP04N2Ha     &     N2                         & 8.562 -0.152 +0.106  & 8.566 -0.150 +0.109  & 8.545 -0.138 +0.095  &   8.475 -0.093 +0.056    \\ 
\hline
\multicolumn{6}{c}{EPOCH 2} \\  
\hline
D02$^{1}$              &     N2                                                     &  8.367     -0.311     +0.162      &   8.331     -0.348     +0.184   &  8.287     -0.311     +0.188   &   8.322     -0.139     +0.119    \\
Z94$^{2}$              &     R$_{23}$                                               &  8.478     -0.032     +0.076      &   8.619     -0.018     +0.030   &  8.571     -0.018     +0.027   &   8.565     -0.012     +0.016     \\
M91$^{3}$              &     R$_{23}$, O3O2 (N2, N2O2)                              &  8.290     -0.038     +0.218      &   8.556     -0.498     +0.045   &  8.142     -0.025     +0.418   &   8.141     -0.021     +0.016     \\
PP04 N2Ha$^{4}$        &     N2                                                     &  8.262     -0.147     +0.104      &   8.240     -0.155     +0.107   &  8.228     -0.156     +0.091   &   8.238     -0.035     +0.026     \\
PP04 O3N2$^{4}$        &     N2, O3/H$\beta$                                        &  8.221     -0.126     +0.067      &   8.210     -0.135     +0.072   &  8.198     -0.134     +0.065   &   8.211     -0.028     +0.019     \\
P10 ONS$^{5}$                &     [N II]$\lambda$6584/H$\beta$, O3/H$\beta$        &  8.558     -0.164     +0.104      &   8.566     -0.165     +0.111   &  8.530     -0.159     +0.103   &   8.520     -0.049     +0.033     \\
                       &     [O II]$\lambda$3727/H$\beta$, [S II]/H$\beta$,   &&&& \\                       
P10 ON$^{5}$                 &     N2, O3/H$\beta$, [O II]$\lambda$3727/H$\beta$    &  7.920     -0.410     +0.264      &   7.957     -0.408     +0.275   &  7.866     -0.393     +0.255   &   7.871     -0.119     +0.079     \\
M08 N2Ha$^{6}$         &     N2                                                     &  8.623     -0.329     +0.067      &   8.541     -0.326     +0.149   &  8.500     -0.325     +0.190   &   8.437     -0.069     +0.047     \\
M08 O3O2$^{6}$         &     O3O2                                                   &  8.654     -0.063     +0.027      &   8.568     -0.037     +0.021   &  8.597     -0.030     +0.019   &   8.625     -0.014     +0.010     \\
M13 O3N2$^{7}$         &     [N II]$\lambda$6584/H$\beta$, O3/H$\beta$              &  8.193     -0.084     +0.044      &   8.185     -0.090     +0.049   &  8.177     -0.089     +0.045   &   8.186     -0.020     +0.015     \\
M13 N2$^{7}$           &     [N II]$\lambda$6584/H$\beta$                           &  8.263     -0.182     +0.098      &   8.234     -0.193     +0.108   &  8.216     -0.193     +0.106   &   8.234     -0.056     +0.049     \\
KD02 N2O2$^{8}$        &     N2O2                                                   &  8.311     -0.741     +0.117      &   8.455     -0.755     +0.118   &  8.366     -0.771     +0.118   &   8.319     -0.763     +0.047     \\
KK04 N2Ha$^{9}$        &     N2, q, (N2O2)                                          &  8.617     -0.146     +0.099      &   8.600     -0.237     +0.138   &  8.564     -0.176     +0.117   &   8.488     -0.122     +0.052     \\
KK04 R23$^{9}$         &     R$_{23}$, q, (N2, N2O2)                                &  8.465     -0.034     +0.165      &   8.691     -0.432     +0.048   &  8.321     -0.020     +0.372   &   8.319     -0.017     +0.013     \\
KD02comb$^{9}$         &     COMBINED$^*$                                           &  8.359     -0.025     +0.022      &   8.173     -0.019     +0.385   &  8.223     -0.018     +0.232   &   8.230     -0.019     +0.015     \\
\hline
KK04         &     R$_{23}$, q, (N2, N2O2)     & 8.352 -0.033 +0.161     &  8.583 -0.463 +0.051     &     8.217 -0.018 +0.333    &   8.215 -0.015 +0.012  \\
Z94          &     R$_{23}$                    & 8.419 -0.043 +0.101     &  8.588 -0.019 +0.032     &     8.534 -0.021 +0.031    &   8.527 -0.014 +0.019  \\
M91          &     R$_{23}$, O3O2 (N2, N2O2)   & 7.837 -0.153 +0.877     &  8.586 -0.907 +0.082     &     7.114 -0.145 +2.429    &   7.108 -0.122 +0.093  \\
D02          &     N2                          & 8.523 -0.249 +0.129     &  8.495 -0.262 +0.138     &     8.463 -0.216 +0.131    &   8.489 -0.103 +0.088  \\
PP04O3N2     &     N2, O3/H$\beta$             & 8.500 -0.107 +0.057     &  8.490 -0.114 +0.061     &     8.480 -0.112 +0.055    &   8.491 -0.024 +0.016  \\
PP04N2Ha     &     N2                          & 8.526 -0.151 +0.107     &  8.504 -0.152 +0.105     &     8.492 -0.149 +0.087    &   8.502 -0.034 +0.025  \\ 
\hline
\multicolumn{6}{c}{EPOCH 3} \\     
\hline
D02$^{1}$              &     N2                                                     &    8.529     -0.197     +0.129    &   8.557     -0.138     +0.106   &    8.545     -0.134     0.108    &   8.408     -0.123     +0.115   \\
Z94$^{2}$              &     R$_{23}$                                               &    8.488     -0.035     +0.081    &   8.621     -0.017     +0.030   &    8.578     -0.017     0.028    &   8.572     -0.010     +0.013   \\
M91$^{3}$              &     R$_{23}$, O3O2 (N2, N2O2)                              &    8.449     -0.030     +0.046    &   8.576     -0.016     +0.026   &    8.539     -0.015     0.023    &   8.532     -0.009     +0.011   \\
PP04 N2Ha$^{4}$        &     N2                                                     &    8.392     -0.124     +0.072    &   8.414     -0.091     +0.057   &    8.405     -0.076     0.049    &   8.294     -0.031     +0.024   \\
PP04 O3N2$^{4}$        &     N2, O3/H$\beta$                                        &    8.309     -0.077     +0.035    &   8.316     -0.051     +0.027   &    8.312     -0.042     0.023    &   8.251     -0.022     +0.016   \\
P10 ONS$^{5}$                &     [N II]$\lambda$6584/H$\beta$, O3/H$\beta$        &    8.706     -0.091     +0.035    &   8.835     -0.062     +0.029   &    8.790     -0.053     0.026    &   8.635     -0.025     +0.017   \\
                       &     [O II]$\lambda$3727/H$\beta$, [S II]/H$\beta$,   &&&& \\           
P10 ON$^{5}$                 &     N2, O3/H$\beta$, [O II]$\lambda$3727/H$\beta$    &    8.440     -0.251     +0.103    &   8.590     -0.172     +0.083   &    8.534     -0.145     0.074    &   8.177     -0.073     +0.050   \\
M08 N2Ha$^{6}$         &     N2                                                     &    8.684     -0.179     +0.071    &   8.699     -0.119     +0.067   &    8.689     -0.102     0.058    &   8.533     -0.052     +0.038   \\
M08 O3O2$^{6}$         &     O3O2                                                   &    8.688     -0.060     +0.024    &   8.565     -0.038     +0.020   &    8.608     -0.030     0.017    &   8.623     -0.014     +0.011   \\
M13 O3N2$^{7}$         &     [N II]$\lambda$6584/H$\beta$, O3/H$\beta$              &    8.251     -0.051     +0.024    &   8.256     -0.034     +0.018   &    8.254     -0.029     0.016    &   8.213     -0.016     +0.012   \\
M13 N2$^{7}$           &     [N II]$\lambda$6584/H$\beta$                           &    8.371     -0.110     +0.059    &   8.388     -0.078     +0.051   &    8.381     -0.065     0.050    &   8.292     -0.044     +0.043   \\
KD02 N2O2$^{8}$        &     N2O2                                                   &    8.600     -0.089     +0.033    &   8.734     -0.049     +0.022   &    8.689     -0.045     0.021    &   8.522     -0.032     +0.021   \\
KK04 N2Ha$^{9}$        &     N2, q, (N2O2)                                          &    8.746     -0.152     +0.064    &   8.856     -0.111     +0.052   &    8.813     -0.094     0.046    &   8.643     -0.063     +0.045   \\
KK04 R23$^{9}$         &     R$_{23}$, q, (N2, N2O2)                                &    8.569     -0.035     +0.052    &   8.711     -0.017     +0.027   &    8.671     -0.017     0.025    &   8.662     -0.010     +0.013   \\
KD02comb$^{9}$         &     COMBINED$^*$                                           &    8.598     -0.109     +0.035    &   8.734     -0.051     +0.022   &    8.689     -0.046     0.021    &   8.522     -0.032     +0.021   \\
\hline
KK04         &     R$_{23}$, q, (N2, N2O2)     &  8.456 -0.036 +0.053     &   8.605 -0.018 +0.029  &  8.562 -0.018+  0.027 & 8.553 -0.011 +0.014   \\
Z94          &     R$_{23}$                    &  8.432 -0.046 +0.106     &   8.590 -0.018 +0.032  &  8.542 -0.019+  0.032 & 8.536 -0.012 +0.015   \\
M91          &     R$_{23}$, O3O2 (N2, N2O2)   &  8.354 -0.077 +0.117     &   8.620 -0.027 +0.044  &  8.553 -0.029+  0.044 & 8.540 -0.018 +0.022   \\
D02          &     N2                          &  8.670 -0.199 +0.130     &   8.699 -0.144 +0.111  &  8.686 -0.138+  0.111 & 8.557 -0.105 +0.098   \\
PP04O3N2     &     N2, O3/H$\beta$             &  8.577 -0.069 +0.031     &   8.583 -0.046 +0.024  &  8.579 -0.038+  0.021 & 8.525 -0.019 +0.014   \\
PP04N2Ha     &     N2                          &  8.672 -0.146 +0.085     &   8.697 -0.108 +0.067  &  8.687 -0.090+  0.058 & 8.560 -0.033 +0.026   \\ 
\hline
Mean Tot[O]+12 (EPOCH 3)        & &8.527 -0.133 +0.135  &8.632 -0.054 +0.054 &8.602 -0.068 +0.068 &8.545 -0.015 +0.016 \\ 
\hline
\end{tabular}

$^*$This method chooses the optimal among given: M91, KD02 N2O2, KD02 N2Ha, KD04 R23, [N2, N2O2] diagnostics \citep{kew08}
References: 
$^1$ Denicol\'{o} et al. 2002        
$^2$ Zaritsky et al. 1994            
$^3$ McGaugh 1991                    
$^4$ Pettini and Pagel 2004          
$^5$ Pilyugin et al. 2010 
$^6$ Maiolino et al. 2008            
$^6$ Marino et al. 2013              
$^7$ Kewley and Dopita 2002          
$^8$ Kobulnicky and Kewley 2004      
$^9$ Kewley and Ellison 2008         
\end{scriptsize}
\end{table*}

\end{appendix}

\end{document}